\documentclass[journal]{IEEEtran}

\usepackage{graphicx,caption,float,subcaption,amssymb,amsmath,color,multirow,mathtools,cite}

\newcounter{mytempeqncnt}

\newcommand{\norm}[1]{\left\lVert#1\right\rVert}
\makeatletter

\newcommand{\Rmnum}[1]{\expandafter\@slowromancap\romannumeral #1@}
\makeatother

\DeclareMathOperator*{\argmax}{arg\,max}

\begin{document}
\title{Interference Mitigation using Optimized Angle Diversity Receiver in LiFi Cellular Network}

\author{Zhihong~Zeng, Chen~Chen,~\IEEEmembership{Member,~IEEE}, Svetislav Savović, Mohammad~Dehghani~Soltani, Cheng~Chen, Majid~Safari,~\IEEEmembership{Senior Member,~IEEE}, Harald~Haas,~\IEEEmembership{Fellow,~IEEE}
\thanks{ This work was financially supported by the National Natural Science Foundation of China under Grant 62271091. (Corresponding author: Chen Chen.)}
\thanks{Zhihong~Zeng and Chen~Chen are with the School of Microelectronics and Communication
	Engineering, Chongqing University, Chongqing, 400000, China. (E-mail: zhihong.zeng@cqu.edu.cn; c.chen@cqu.edu.cn.)}
\thanks{Svetislav Savović is with University of Kragujevac, Faculty of Science, R. Domanovića 12, Kragujevac, Serbia. (E-mail:	savovic@kg.ac.rs).}
\thanks{Mohammad Dehghani Soltani and Majid Safari are with the School of Engineering, Institute for Digital Communications, The University of Edinburgh, Edinburgh EH9 3FD, U.K. (E-mail: m.dehghani@ed.ac.uk; majid.safari@ed.ac.uk).}
\thanks{Cheng~Chen and Harald Haas are with the	LiFi Research and Development Centre (LRDC), Department of Electronic and Electrical Engineering, University of Strathclyde, Glasgow G1 1RD, U.K. (E-mail: c.chen@strath.ac.uk; harald.haas@strath.ac.uk).}
}

\maketitle

\begin{abstract}
Light-fidelity (LiFi) is an emerging technology for high-speed short-range mobile communications. Inter-cell interference (ICI) is an important issue that limits the system performance in an optical attocell network. Angle diversity receivers (ADRs) have been proposed to mitigate ICI. In this paper, the structure of pyramid receivers (PRs) and truncated pyramid receivers (TPRs) are studied. The coverage problems of PRs and TPRs are defined and investigated, and the lower bound of field of view (FOV) for each PD is given analytically. The impact of random device orientation and diffuse link signal propagation are taken into consideration. The performances of PRs and TPRs are compared and then optimized ADR structures are proposed. The performance comparison between the select best combining (SBC) and maximum ratio combining (MRC) is given under different noise levels. It is shown that SBC will outperform MRC in an interference limited system, otherwise, MRC is a preferred scheme. In addition, the double source system, where each LiFi AP consists of two sources transmitting the same information signals but with opposite polarity, is proved to outperform the single source system under certain conditions.
\end{abstract}



\section{Introduction}
Due to the increasing demand for wireless data, the radio frequency (RF) spectrum has become a very limited resource. Alternative approaches are investigated to support future growth in data traffic and next-generation high-speed wireless communication systems. Techniques such as massive multiple-input multiple-output (MIMO), millimeter wave (mmWave) communications  and Light-Fidelity (LiFi) are being explored. Among these technologies, LiFi is a novel bi-directional, high-speed and fully networked wireless communication technology. A typical LiFi system uses off-the-shelf low-cost commercially available light emitting diodes (LEDs) and photodiodes (PDs) as front end devices \cite{OBrienD}. Intensity modulation (IM) is used to encode the information in visible light communication (VLC) since the LED is an incoherent optical source. Direct detection (DD) is adopted at the receiver end. LiFi utilizes visible light as the propagation medium in the downlink for both illumination and communication purposes. It may use infrared light in the uplink in order to allow the illumination constraint of the room to be unaffected, and also to avoid interference with the visible light in the downlink \cite{Haas}. The overall license-free bandwidth of visible light is more than $1000$ times greater than the whole RF spectrum \cite{Haas}. Also, LiFi can provide enhanced security as the light does not penetrate through opaque objects \cite{WuVLC5G}. In many large indoor environments, multiple light fixtures are installed, these luminaries can act as VLC access points (APs). A network consisting of multiple VLC APs is referred to as a LiFi attocell network \cite{Haas}. Given the widespread use of LED lighting, a LiFi attocell network can use the existing lighting infrastructures to offer fully networked wireless connectivity. Moreover, LiFi attocells can be regarded as an additional network layer within the existing wireless networks because there is no interference to the RF counterparts such as femtocell networks \cite{Haas}. These benefits of LiFi have made it favorable for recent and future research.	

By improving the spatial reuse of the spectrum resources, cellular networks can achieve a higher area spectral efficiency \cite{Alouini}. In comparison with RF femtocell networks, LiFi attocell networks use smaller cell sizes as the light beams from LEDs are intrinsically narrow \cite{SpotlightingforVLC}. Thus, with the densely deployed optical APs, the LiFi attocell network can achieve a better bandwidth reuse and a higher area spectral efficiency. However, similar to other cellular systems, inter-cell interference (ICI) in LiFi attocell networks limits the system performance. This is because the signal transmitted to a user will interfere with other users who are receiving signals from the same frequency resource. Particularly, cell-edge users suffer from severe ICI. Despite the dense deployment of APs, due to ICI, LiFi may not provide a uniform coverage concerning data rate. Interference coordination mechanisms have been extensively investigated for VLC systems \cite{Marsh,Cui,Chen,JointTrans_GlobCom_Cheng,ADT_JLT_Zhe}. The commonly used technique is static resource partitioning \cite{Marsh}. By separating any two cells that reuse the same frequency resource with a minimum reuse distance, ICI is effectively mitigated. However, there is a significant loss in spectral efficiency. A combined wavelength division and code division multiple access scheme was proposed in \cite{Cui}. Although this approach enhances the system bandwidth, it requires separate filters for each color band and thus creates additional cost. In \cite{Chen}, the fractional frequency reuse (FFR) technique is proposed to mitigate ICI. The FFR scheme is a cost-effective approach to provide improvements both in cell-edge user performance and average spectral efficiency, but a low user-density will decrease the average spectral efficiency significantly. Joint transmission (JT) has been proven to improve signal quality for cell-edge users \cite{JointTrans_GlobCom_Cheng}. The downside of the JT systems is the extra signaling overhead. Moreover, the space division multiple access (SDMA) scheme using angle diversity transmitters proposed in \cite{ADT_JLT_Zhe} can mitigate ICI by generating concentrated beams to users at different locations.

The angle diversity reception, first proposed in \cite{KJM}, allows the receiver to achieve a wide field of view (FOV) and high optical gain simultaneously. An angle diversity receiver (ADR) is composed of multiple narrow-FOV PDs facing in different directions. In \cite{ADR_NLOS_ICC,ADR_NLOS_TCOM,FlyEye_1992,ADR_JLT_Zhe,ADR_VTC_Zhe,DoubSource_GlobCom_Zhe,ADR_ICC_ZENG,ADR_WCNC_ZENG}, the ADR is used to address the issue of ICI as well as frequency reuse in LiFi cellular systems, and different signal combining schemes are investigated. However, the proposed ADR structure is hard to implement and the optimum ADR design is not given. Moreover, the system is assumed to be interference limited instead of noise limited in \cite{ADR_JLT_Zhe}, which is not always true as the ADR can mitigate most of ICI with noise being the dominated part. Recently, due to the lower channel correlation achieved from the angle diversity scheme, ADRs are introduced to improve the performance of indoor MIMO-VLC systems, and the pyramid receivers (PRs) are proposed \cite{PR_ADR_MIMO}.  The generalized structure of truncated pyramid receivers (TPRs) are given in \cite{TPR_ADR} to reduce the signal to interference plus noise ratio (SINR) fluctuation. However, the optimum structures of the ADRs are not given and therefore the performance gain is not fully exploited. 
In addition, to obtain a more accurate evaluation of the system performance, the following three factors must be taken into consideration:
\subsubsection{User Device Orientation} Most of the studies on ADRs assume that the receiving device is pointed vertically upward. However, it has been shown in our previous works that the random orientation of mobile devices can significantly affect the direct current (DC) channel gain and thus the system performance \cite{MDSorientation,ZhihongOrientation}. Therefore, the random orientation of the user equipment (UE) needs to be considered. A random device orientation model has been proposed in \cite{MDSorientation}. This model will be applied in this study to evaluate the system more accurately. 
\subsubsection{Diffuse Link Signal Propagation} The non-line-of-sight (NLOS) link is neglected in most LiFi and VLC studies and only the line-of-sight (LOS) channel is considered \cite{Alouini,SpotlightingforVLC,Marsh,Cui,Chen,JointTrans_GlobCom_Cheng}. In \cite{Downlink_Performance_OWC}, it is shown that the LOS link is the dominate link and the effect of the reflected signal can be neglected. However, the UE is assumed to be positioned vertically upward, which is not realistic for mobile devices. In our study, we consider the effect of reflection when random device orientation is applied and the results show that the diffuse link cannot be ignored. A microscopic frequency-domain method for the simulation of the indoor VLC channel is presented in \cite{FrequencyDomain2016}. A closed form for the transfer function that contains all reflection orders is formulated. The method can be extended to multi-spot transmission without a significant increase in the computational complexity. Therefore, in this study, we will use the frequency-domain method to simulate the impact of the diffuse link.
\subsubsection{Noise Power Spectral Density}
The noise power spectral density of the PD has a huge impact on the analyses of system performance. For different levels of noise power spectral density, the system could be noise-limited, interference-limited or noise-plus-interference limited, which could affect the choices of the signal combing schemes and the cell configurations.

The main contributions of this paper are summarized as follows:
\begin{itemize}
	\item The coverage area of ADRs is defined to differentiate from the coverage area of APs. Analytical expressions for the coverage area of both PRs and TPRs are given.
	\item Based on the constraint set by the coverage area of ADRs, the lower bound of FOV of PDs on an ADR is given for the single source (SS) system. The performance of PRs and TPRs are compared, and optimized ADR structures are proposed to fully exploit the performance gain of ADRs. In addition, the joint effect of the receiver and transmitter bandwidth on the average data rate are analyzed.
	\item The performance comparison between the select best combining (SBC) and maximum ratio combining (MRC) are given regarding different levels of noise power spectral density. It is the first time shown that under certain circumstances, the SBC can outperform the MRC. 
	\item The double source (DS) cell system is considered to further mitigate the NLOS interference. The lower bound of FOV of PDs on an ADR is derived and the optimized ADR structures are proposed for the DS system. 
	\item  By comparing the average SINR between the DS system and the SS system under different levels of noise power spectral density, we present that, in a noise-dominated scenario, the SS system should be applied, otherwise, the DS system is preferred. 
\end{itemize}

The rest of this paper is organized as follows. The system model is introduced in Section $\rm{\Rmnum{2}}$. The generalized structures of ADRs are given in Section $\rm{\Rmnum{3}}$. Section $\rm{\Rmnum{4}}$ presents the optimum FOV for PRs and TPRs. The concepts of the optical double-source cell are proposed in Section $\rm{\Rmnum{5}}$. The simulation results and discussions are presented in Section $\rm{\Rmnum{6}}$. Finally, conclusions are drawn in Section $\rm{\Rmnum{7}}$.

\section{System Model}

\subsection{Light Propagation Model}
In indoor optical communications, the signal propagation consists of two components: the LOS link and the diffuse link. 


\subsubsection{LOS link}
It is typically assumed that the LED follows the Lambertian radiation pattern and the the LOS DC channel gain between the transmitter (Tx) and receiver (Rx) is given by \cite{KJM,Haas}:
\begin{equation}
	H_{\rm{LOS}}=\frac{(m+1)}{2{\pi}d^2}A_{\rm{p}}T_s(\psi)g(\psi)\cos^m(\phi)\cos(\psi)v_{\rm{Tx},\rm{Rx}},	
	\label{eq_hlos}
\end{equation}
where $m$ is the Lambertian order, which is given as $m=-\ln(2)/\ln(\cos(\Phi_{1/2}))$, and $ \Phi_{1/2} $ denotes the half-power semi-angle of the LED; $d$ is the distance between the Tx and the Rx; $A_{\rm{p}}$ denotes the physical area of the PD; $T_s(\psi)$ represents the signal transmission gain of the optical filter; The irradiance angle of the transmitter is denoted as $\phi$ and the incidence angle of the receiving PD is denoted as $\psi$. Note that $\psi$ can be obtained by $\cos(\psi)=\frac{\mathbf{n}_{\rm{PD}}\cdot \mathbf{d}}{\lVert \mathbf{d} \rVert}$, where $\mathbf{d}$ defines the distance vector between the Tx and the Rx. The dot product is denoted as $(\cdot)$ and $\lVert \mathbf{d} \rVert$ denotes the Euclidean distance. Furthermore, $\mathbf{n}_{\rm{PD}}$ is the normal vector of the PD. $g(\psi)$ is the concentrator gain and \mbox{$g(\psi)=\frac{n^2_{\rm{ref}}}{\sin^2(\Psi_{\rm{c}})}$}, where $n_{\rm{ref}}$ represents the internal refractive index of the concentrator and $\Psi_{\rm{c}}$ denotes the FOV of the PD with concentrator. $v_{\rm{Tx},\rm{Rx}}$ represents the visibility factor and is given by \cite{FrequencyDomain2016}: 
\begin{equation}
	v_{\rm{Tx},\rm{Rx}}=\left\{
	\begin{array}{ll}
		0,	 &  \phi>\pi/2 \ \text{or} \ \psi>\Psi_{\rm{c}} \\
		1 &  \ \text{otherwise}
	\end{array}
	\right..	
	\label{eq_v}
\end{equation}

\subsubsection{NLOS link}
The diffuse link is due to the reflection from the walls. As mentioned earlier, the frequency-domain method in \cite{FrequencyDomain2016} is used to obtain the diffuse link DC channel gain. We assume that all the wall surfaces are purely diffuse Lambertian reflectors with $m=1$. All of the surfaces are divided into a number of small surface elements numbered by $k=1,...,N_{\rm{E}}$, with areas $A_k$ and reflective coefficients $\rho_k$. To calculate the diffuse link DC channel gain, the propagation of light is divided into the following three parts. The first part of the diffuse link propagation is the light path between the $\rm{Tx}$ and all the reflective surface elements of the room. The LOS DC channel gain between the $\rm{Tx}$ and the surface element $k$ is defined as $H_{\rm{Tx},k}$.
The transmitter transfer vector, $\mathbf{t}$, is defined as	$\mathbf{t}=(H_{\rm{Tx,1}}, \ H_{\rm{Tx,2}}, \ ... \ H_{\rm{Tx},N_{\rm{E}}})^{\rm{T}}$, where $(\cdot)^T$ defines the transpose of vectors.
The second part of the diffuse link is the LOS link from all the $N_{\rm{E}}$ surface elements to all the $N_{\rm{E}}$ surface elements. The LOS DC channel gain between the surface elements $k$ and the surface element $i$ is given as $H_{k,i}$.
To describe the LOS links between all surfaces inside the room, the $N_{\rm{E}} \times N_{\rm{E}}$ room-intrinsic transfer matrix, ${\mathbf{H}}$, is defined by its elements $[{\mathbf{H}}]_{k,i}=H_{k,i}$. In order to include the reflective coefficient $\rho_k$ of the surface elements, the $N_{\rm{E}} \times N_{\rm{E}}$ reflectivity matrix is defined as $\mathbf{G}_{\rho}={\rm{diag}}(\rho_1,\rho_2,...,\rho_{N_{\rm{E}}})$ \cite{FrequencyDomain2016}.
In the third part of the diffuse link, the light propagates from all the surfaces of the room to the $\rm{Rx}$. Similarly, we denote the LOS DC channel gain between the surface element $k$ and the $\rm{Rx}$ as $H_{k,{\rm{Rx}}}$.
The LOS DC channel gain between all the reflective elements of the room and the receiver are grouped to give the receiver transfer vector $\mathbf{r}$ which is defined by its transpose $\mathbf{r}^{\rm{T}}=(H_{1,\rm{Rx}}, \ H_{2,\rm{Rx}}, \ ... \ H_{N_{\rm{E}},{\rm{Rx}}})$. Therefore, the total diffuse DC channel gain with infinite reflection can be calculated by the matrix product \cite{FrequencyDomain2016}:
\begin{equation}
	\label{eq_total_Transfer_func}
	H_{\rm{diff}}=\mathbf{r}^{\rm{T}}\mathbf{G}_{\rho}(\mathbf{I}-{\mathbf{H}}\mathbf{G}_{\rho})^{-1}\mathbf{t},
\end{equation}
where $\mathbf{I}$ denotes the unity matrix. 
\subsection{Signal Combining Schemes for ADR} 

An indoor LiFi network is studied  and it is assumed that the total number of UE and LiFi APs are $N_{\rm{UE}}$ and $N_{\rm{L}}$, respectively. The set of APs is denoted by $\mathcal{A}=\{\alpha \mid \alpha \in[1,N_{\rm{L}}] \}$. The set of users is denoted as $\mathcal{U}=\{\mu\mid\mu\in[1,N_{\rm{UE}}] \}$.  The ADR is used as the Rx and the set of PDs on an ADR is denoted as $\mathcal{P}=\{p\mid p\in[1,N_{\rm{PD}}] \}$, where $N_{\rm{PD}}$ denotes the total number of PDs on the ADR.
As one of the most commonly used optical orthogonal frequency division multiplexing (O-OFDM) schemes, the direct current biased optical OFDM (DCO)-OFDM is used in this study as it is spectrum efficient \cite{OSS}. The number of OFDM subcarriers is denoted as $M$, where $M$ is an even and positive integer, and the sequence number of OFDM subcarriers is denoted by $m \in \{0,1, \dots, M-1\}$. Two constraints should be satisfied to ensure real and positive signals: i) $X(0) = X(M/2) = 0$, and ii) the Hermitian symmetry constraint, i.e., $X(m) = X^*(M-m)$, for $m\neq0$, where $(\cdot)^*$ denotes the complex conjugate operator  \cite{DPO}. Therefore, the effective subcarrier set bearing information data is defined as $\mathcal{M}_{\rm e} = \{m| m \in [1, M/2-1], m \in \mathbb{N}\}$, where $\mathbb{N}$ is the set of natural numbers.

\begin{figure*}[!t]
	\normalsize
	\setcounter{mytempeqncnt}{\value{equation}}
	\setcounter{equation}{10}
	\begin{equation}
		\centering
		(x_{{\rm{PD}},p}, y_{{\rm{PD}},p}, z_{{\rm{PD}},p})= 
		\begin{dcases}
			\Big(x_{\rm{UE}}+r\cos\frac{2(p-1)\pi}{N_{\rm{TPR}}-1}, \ y_{\rm{UE}}+r\sin\frac{2(p-1)\pi}{N_{\rm{TPR}}-1}, \ z_{\rm{UE}}\Big), & \text{if} \ 1 \leq p < N_{\rm{TPR}}\\
			(x_{\rm{UE}}, \ y_{\rm{UE}}, \ z_{\rm{UE}}),& \text{if} \ p=N_{\rm{TPR}}
		\end{dcases}
		\label{eq_coordinate_PD}
	\end{equation}
	\setcounter{equation}{\value{mytempeqncnt}}
	
	\hrulefill
	\vspace*{4pt}
\end{figure*}

For an ADR, multiple PDs are receiving signals simultaneously. Thus, attention should be paid to the selection of the signal combing schemes. There are different combining schemes such as equal gain combining (EGC), SBC and MRC. An important metric to evaluate the link quality and capacity is the SINR. The SINR of user $\mu$ on subcarrier $m$ can be obtained based on \cite{ADR_JLT_Zhe} and \cite{DownlinkPerformanceChen}:
\begin{equation}
	\resizebox{0.9\hsize}{!}{$
		\centering
		\begin{split}
			\gamma_{\mu,m}&=\frac{(\sum\limits_{p=1}^{N_{\rm{PD}}}\tau P_{\rm{tx}}w_pH_{\alpha_{\rm{s}},\mu,p})^2/(M-2)}{\sum\limits_{p=1}^{N_{\rm{PD}}}{w_p}^2\kappa^2N_0B_{\rm{L}}/M+\sum\limits_{\alpha_{\rm{i}}\in\mathcal{A}\setminus \{\alpha_{\rm{s}}\}}(\tau P_{\rm{tx}}\sum\limits_{p=1}^{N_{\rm{PD}}}w_pH_{\alpha_{\rm{i}},\mu,p})^2/(M-2)}\\
			&=\frac{(\sum\limits_{p=1}^{N_{\rm{PD}}}\tau P_{\rm{tx}}w_pH_{\alpha_{\rm{s}},\mu,p})^2}{\sum\limits_{p=1}^{N_{\rm{PD}}}{w_p}^2\kappa^2N_0B_{\rm{L}}(M-2)/M+\sum\limits_{\alpha_{\rm{i}}\in\mathcal{A}\setminus \{\alpha_{\rm{s}}\}}(\tau P_{\rm{tx}}\sum\limits_{p=1}^{N_{\rm{PD}}}w_pH_{\alpha_{\rm{i}},\mu,p})^2}
		\end{split} ,
		\label{eq_sinr}
		$}
\end{equation}
where $\tau$ is the optical-to-electrical conversion efficiency; $P_{\rm{tx}}$ is the transmitted optical power of the AP;  $w_p$ denotes the combining weight of PD $p$; $H_{\alpha_{\rm{s}},\mu,p}$ is the overall DC channel gain between the PD $p$ of user $\mu$ and the serving AP $\alpha_{\rm{s}}$; $\kappa$ is the ratio of DC optical power to the square root of electrical signal power; $N_0$ represents the noise power spectral density of the additive white Gaussian noise and $B_{\rm{L}}$ is the baseband modulation bandwidth;  $H_{\alpha_{\rm{i}},\mu,p}$ is the overall DC channel gain between the PD $p$ of user $\mu$ and the interfering LiFi AP $\alpha_{\rm{i}}$. The serving AP $a_{\rm{s}}$ for user $\mu$ is selected based on the signal strength strategy (SSS) where the UEs are connected to the APs providing the best received signal strength. Hence, the serving AP $\alpha_{\rm{s}}$ for user $\mu$ can be expressed as:
\begin{equation}
	\alpha_{\rm{s}}=\argmax\limits_{\alpha\in\mathcal{A}} \sum\limits_{p=1}^{N_{\rm{PD}}}|H_{\alpha,\mu,p}|^2.
	\label{eq_servingAP}
\end{equation}
When the EGC scheme is adopted, the signals received by the PDs are simply combined with equal weights, which can be described as:
\begin{equation}
	w_p=1, \ \ {\text{for any }} p\in \mathcal{P}.
	\label{eq_EGC}
\end{equation}

In terms of the SBC scheme, a switch circuit is required to output the information from the PD with the highest SINR. Hence, the weight of each PD is given as:
\begin{equation}
	\setcounter{equation}{7}
	w_p=
	\begin{cases}
		1, & p=p_{\rm{s}}\\
		0, & {\rm{otherwise}}
	\end{cases},
	\label{eq_SBC}
\end{equation}
where $p_{\rm{s}}$ can be obtained by:
\begin{equation}
		\resizebox{\hsize}{!}{$
	p_{\rm{s}}=\argmax\limits_{p\in\mathcal{P}}\frac{(\tau P_{\rm{tx}}H_{\alpha_{\rm{s}},\mu,p})^2}{\kappa^2N_0B_{\rm{L}}(M-2)/M+\sum\limits_{\alpha_{\rm{i}}\in\mathcal{A}\setminus \{\alpha_{\rm{s}}\}}(\tau P_{\rm{tx}}H_{\alpha_{\rm{i}},\mu,p})^2}.
	$}
\end{equation}
On the subject of the MRC schemes, 
the weight for each PD is denoted as \cite{ADR_JLT_Zhe}:
\begin{equation}
	w_p=\frac{(\tau P_{\rm{tx}}H_{\alpha_{\rm{s}},\mu,p})^2}{\kappa^2N_0B_{\rm{L}}(M-2)/M+\sum\limits_{\alpha_{\rm{i}}\in\mathcal{A}\setminus \{\alpha_{\rm{s}}\}}(\tau P_{\rm{tx}}H_{\alpha_{\rm{i}},\mu,p})^2}
	\label{eq_MRC}.
\end{equation}
Based on the Shannon capacity, assuming electrical signals after optical to electrical conversion, the data rate of the $\mu$-th UE on subcarrier $m$ can be expressed as \cite{Dimitrov2013}:
\begin{equation}
	\zeta_{\mu,m}=
	\begin{cases}
		\frac{B_{\rm{L}}}{M} \log_2(1+\gamma_{\mu,m}), &  m \in [1, M/2-1]\\
		0, & {\rm{otherwise}}
	\end{cases}
	\label{eq_SE}.
\end{equation}
Hence, the data rate of the $\mu$-th UE can be obtained by $\zeta_{\mu}=\sum_{m=1}^{M/2-1}\zeta_{\mu,m}$.

\section{ADR Structure}
\begin{figure}[!b]
	\centering
	\vspace{-20pt}
	\begin{subfigure}[h]{\columnwidth}
		\centering
		\includegraphics[height=0.7\columnwidth]{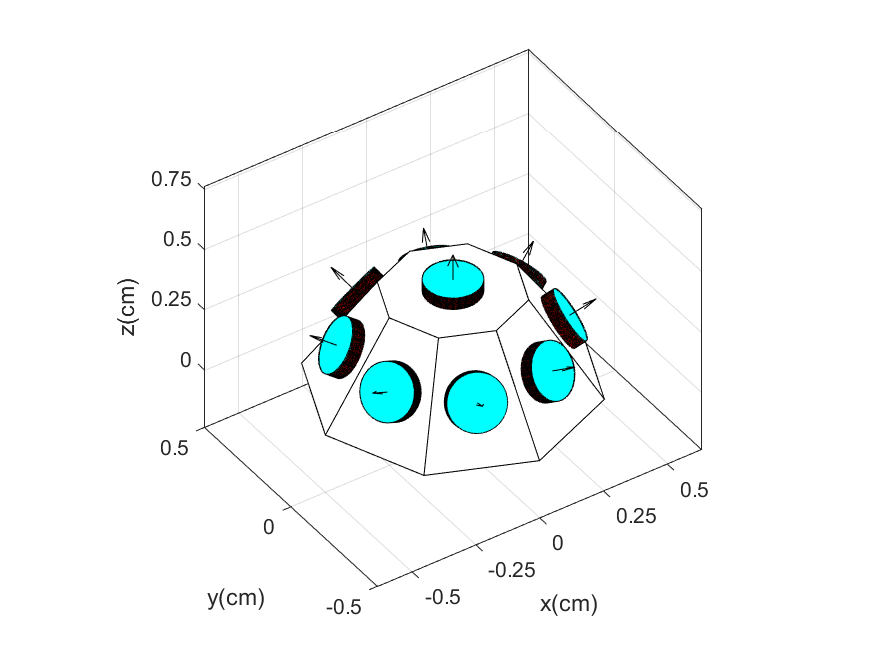}
		\caption{The structure of TPR with $N_{\rm{TPR}}=9$.}
		\label{fig_TPR} 
	\end{subfigure}%
	\\
	\begin{subfigure}[h]{\columnwidth}
		\centering
		\includegraphics[height=0.7\columnwidth]{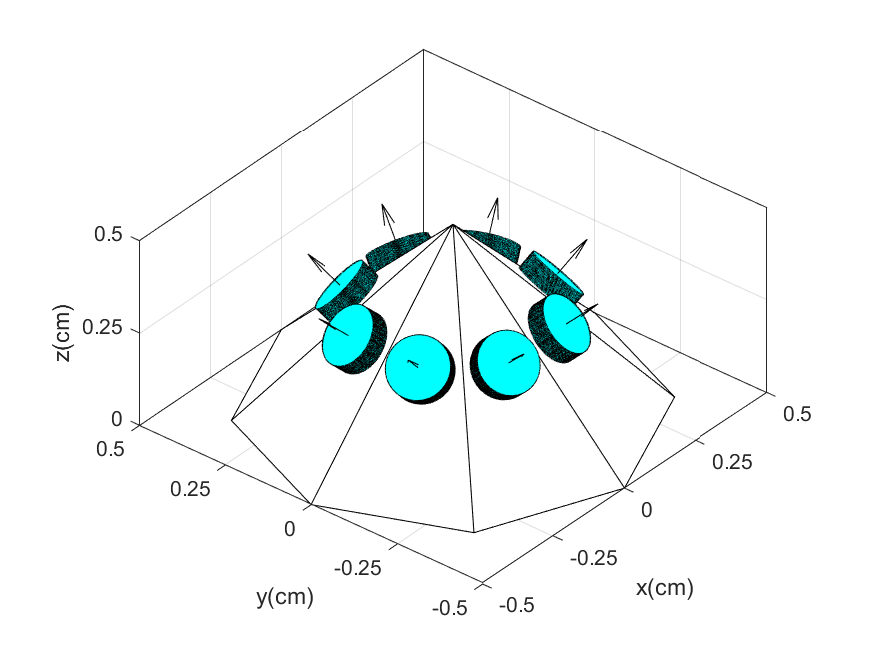}
		\caption{The structure of PR with $N_{\rm{PR}}=8$.}
		\label{fig_PR}
	\end{subfigure}
	\caption{ADR structures.}
\end{figure}

The ADR is composed of multiple PDs facing in different directions. By using a PD in conjunction with a compound parabolic concentrator (CPC), a narrow FOV and high optical gain can be achieved \cite{KJM}. However, the narrow FOV is achieved at the expense of the longer length of the CPC. Therefore, the number of PDs on the ADR should be limited due to the size limitation on the mobile devices and smartphones. In this study, the TPR \cite{TPR_ADR} and the PR \cite{PR_ADR_MIMO} are considered as  they are both suitable for hand-held devices. The number of PDs on the TPR and  PR are separately denoted as $N_{\rm{TPR}}$ and $N_{\rm{PR}}$.  The structure of the TPR with $N_{\rm{TPR}}=9$ and the PR with $N_{\rm{PR}}=8$ are presented in Fig. \ref{fig_TPR} and \ref{fig_PR}, respectively. The ADR designs are analyzed in the following parts.

\begin{figure*}[!t]
	\normalsize
	\setcounter{mytempeqncnt}{\value{equation}}
	
	\setcounter{equation}{16}
	\begin{equation}
		\begin{split}
			\mathbf{n}_{\rm{UE}} = \mathbf{R})(\omega_{\rm{UE}} \mathbf{R}(\theta_{\rm{UE}})  \mathbf{n}_{\rm{UE,vert}} 
			&=
			[\sin\theta_{\rm{UE}}\cos\omega_{\rm{UE}} ,	\sin\theta_{\rm{UE}}\sin\omega_{\rm{UE}},	\cos\theta_{\rm{UE}} ]^{\rm{T}}
		\end{split}
		\label{eq_rotation}
	\end{equation}
	
	\setcounter{equation}{20}
	\begin{equation}
		\omega_{{\rm{PD}},p}=\tan^{-1}\Big( \frac{	C_1\sin\omega_{\rm{UE}}\cos\theta_{\rm{UE}} + C_2\cos\omega_{\rm{UE}}  + \sin\theta_{\rm{UE}}\sin\omega_{\rm{UE}} \cos (\theta_{{\rm{PD,vert}},p})}{C_1\cos\omega_{\rm{UE}}\cos\theta_{\rm{UE}} -C_2\sin\omega_{\rm{UE}}  + \sin\theta_{\rm{UE}}\cos\omega_{\rm{UE}}\cos (\theta_{{\rm{PD,vert}},p})} \Big)
		\label{eq_omege_pd_p}
	\end{equation} 
	
	\setcounter{equation}{\value{mytempeqncnt}}
	\hrulefill
	\vspace*{4pt}
\end{figure*}

\subsection{TPR Design}
The TPR is composed of a central PD and a ring of $N_{\rm{TPR}}-1$ equally separated side PDs. The side PDs are arranged uniformly in a circle of radius $r$ on the horizontal plane. Thus, the coordinate of the $p$-th PD on a TPR is represented as \eqref{eq_coordinate_PD}, where $(x_{\rm{UE}},y_{\rm{UE}},z_{\rm{UE}})$ is the UE position, denoted as $\mathbf{p}_{\rm{UE}}$ \cite{ADR_WCNC_ZENG}. As the distance between the AP and the UE is much larger than $r$, the distances between the AP and all PDs on a TPR are approximately the same. The normal vector of each PD can be described by two angles: the azimuth angle of a PD, $\omega_{\rm{PD}}$, and the elevation angle of a PD, $\theta_{\rm{PD}}$ \cite{ADR_ICC_ZENG}.
When the UE is pointing vertically upward, the TPR has one vertically orientated central PD and $N_{\rm{TPR}}-1$ inclined side PDs with identical elevation angles $\Theta_{\rm{PD}}$. In other words, the elevation angle of the $p$-th PD on a TPR can be expressed as:
\begin{equation}
	\setcounter{equation}{12}
	\centering
	\theta_{{\rm{PD,vert}},p}= 
	\begin{dcases}
		\Theta_{\rm{PD}}, & \text{if} \ 1 \leq p < N_{\rm{TPR}}\\
		0, & \text{if} \ p=N_{\rm{TPR}}
	\end{dcases}.
\end{equation}
The azimuth angle of the $p$-th PD is given by:
\begin{equation}
	\centering
	\omega_{{\rm{PD,vert}},p}= 
	\begin{dcases}
		\frac{2(p-1)\pi}{N_{\rm{TPR}}-1}, & \text{if} \ 1 \leq p < N_{\rm{TPR}}\\
		0, & \text{if} \ p=N_{\rm{TPR}}
	\end{dcases}.
\end{equation}

\subsection{PR Design}
The PR can be regarded as a TPR without the central PD. Therefore, the coordinate of the $p$-th PD on a PR is given by: 
\begin{equation}
		\resizebox{0.8\hsize}{!}{$
			\begin{split}
				&(x_{{\rm{PD}},p}, y_{{\rm{PD}},p}, z_{{\rm{PD}},p})=\\ &\Big(x_{\rm{UE}}+r\cos\frac{2(p-1)\pi}{N_{\rm{PR}}}, \ y_{\rm{UE}}+r\sin\frac{2(p-1)\pi}{N_{\rm{PR}}}, \ z_{\rm{UE}}\Big).
			\end{split}	
		$}
\end{equation}
When the UE is vertically orientated, the elevation angle and the azimuth angle of the $p$-th PD are separately expressed as:
\begin{equation}
	\theta_{{\rm{PD,vert}},p}= \Theta_{\rm{PD}}, \ \ \ \  
	\omega_{{\rm{PD,vert}},p}=\frac{2(p-1)\pi}{N_{\rm{PR}}}.
\end{equation}

\subsection{Random Orientation Model}
\label{Section_Random_Orientation}
\begin{figure}[!b]
	\centering
	\includegraphics[width=0.5\columnwidth]{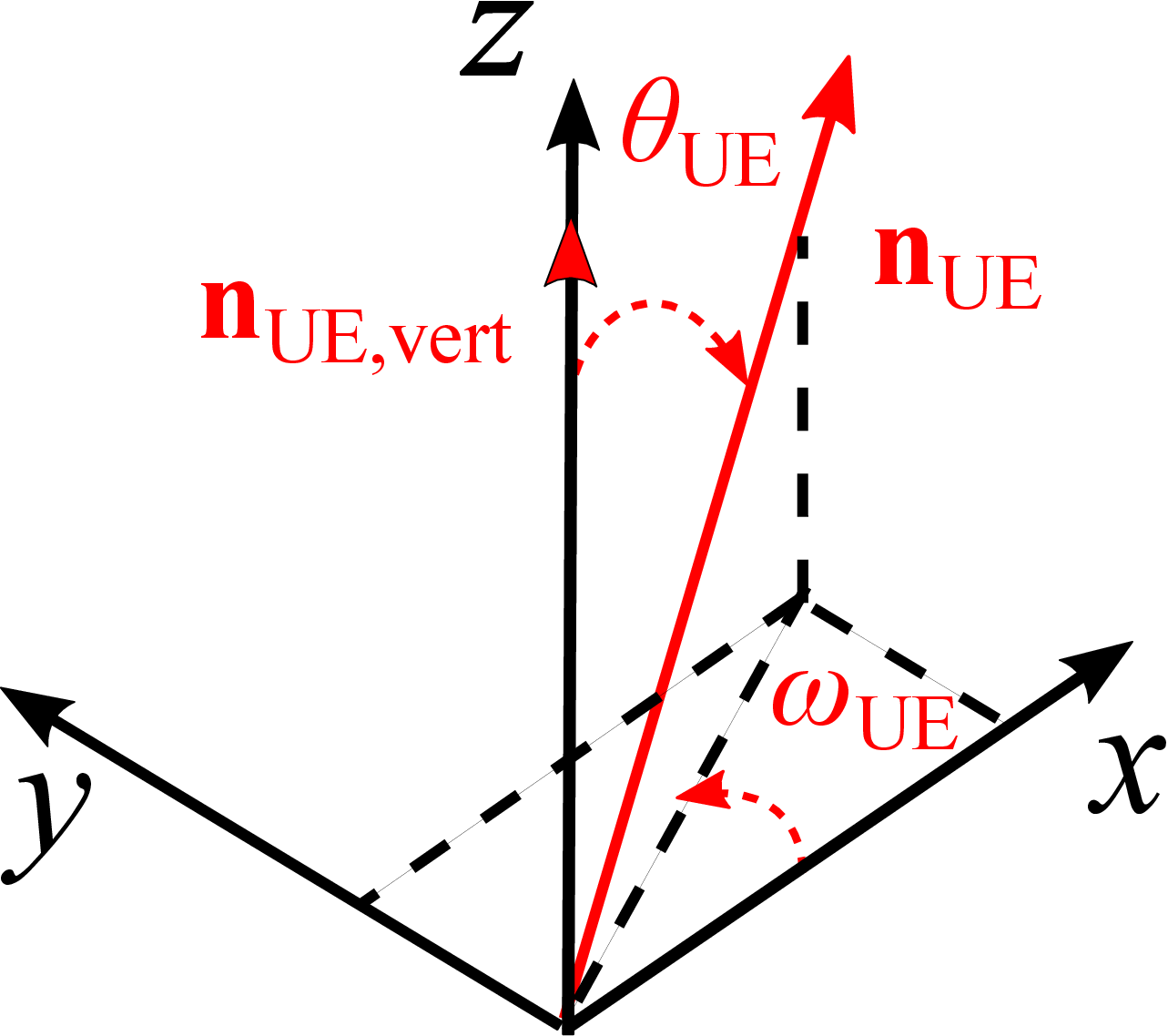}
	\caption{Representation of random UE orientation.}
	\label{fig_rotation}
\end{figure}
The orientation of a UE has a great impact on the channel DC gain according to \eqref{eq_hlos}. In \cite{MDSorientation}, a model for the random orientation of mobile devices based on experiments is proposed so that the system performance of LiFi attocell networks can be evaluated more accurately. The random orientation model can be described by two angles: the elevation angle of a UE, $\theta_{\rm{UE}}$, and the azimuth angle of a UE, $\omega_{\rm{UE}}$. The geometrical representation of $\theta_{\rm{UE}}$ and $\omega_{\rm{UE}}$ is manifested in Fig. \ref{fig_rotation}. The probability density function (PDF) of $\theta_{\rm{UE}}$ can be modeled as the truncated Laplace distribution and it can be simplified as \cite{MDSorientation}:
\begin{equation}
	f_{\theta}(\theta_{\rm{UE}}) \cong \frac{\exp(-\frac{|\theta_{\rm{UE}}-\mu_{\theta}|}{b_{\theta}})}{2b_{\theta}}, \ \ \  0 \leq \theta \leq \frac{\pi}{2},
\end{equation}
where $b_{\theta}=\sqrt{\sigma^2_{\theta}/2}$. The mean and scale parameters are set as $\mu_{\theta}=41.39^\circ$ and $\sigma_{\theta}=7.68^\circ$ \cite{MDSorientation}.
In addition, the PDF of the azimuth angle of a UE, $\omega_{\rm{UE}}$, is modeled as a uniform distribution.  It is assumed that the UE is initially pointing vertically upward and $\mathbf{n}_{\rm{UE,vert}}=[0, 0, 1]^{\rm{T}}$. The normal vector of the UE after rotation becomes $\mathbf{n}_{\rm{UE}}$. The rotation can be simplified as rotating around the y-axis with $\theta_{\rm{UE}}$ and then rotating around z-axis with $\omega_{\rm{UE}}$, which can be described by rotation matrices $\mathbf{R}(\theta_{\rm{UE}})$ and $\mathbf{R}(\omega_{\rm{UE}})$ separately \cite{LSM}. Thus, $\mathbf{n}_{\rm{UE}}$ is given by \eqref{eq_rotation}.

\subsection{Normal Vector of the ADR}
When the UE is pointing vertically upward, for both PRs and TPRs, the normal vector of the $p$-th PD is obtained as:
\begin{equation}
	\setcounter{equation}{18}
	\begin{aligned}
		\mathbf{n}_{{\rm{PD,vert}},p}=		\left[ {\begin{array}{c}
				\sin (\theta_{{\rm{PD,vert}},p}) \cos (\omega_{{\rm{PD,vert}},p}) \\
				\sin (\theta_{{\rm{PD,vert}},p})\sin (\omega_{{\rm{PD,vert}},p})\\
				\cos (\theta_{{\rm{PD,vert}},p})\\
		\end{array} } \right].
	\end{aligned}
	\label{eq_normalV}
\end{equation}
However, the normal vector of the UE will change due to the random rotation. The random orientation model is described in Section \ref{Section_Random_Orientation}. 
Thus, the normal vector of the $p$-th PD after the random rotation is obtained by:
\begin{equation}
	\resizebox{\hsize}{!}{$
		\begin{aligned}
			&\mathbf{n}_{{\rm{PD}},p}(\theta_{\rm{UE}},\omega_{\rm{UE}})
			=\mathbf{R}(\omega_{\rm{UE}}) \mathbf{R}(\theta_{\rm{UE}})  \mathbf{n}_{{\rm{PD,vert}},p}\\
			&=		\left[ {\begin{array}{c}
					C_1\cos\omega_{\rm{UE}}\cos\theta_{\rm{UE}} -C_2\sin\omega_{\rm{UE}}  + \sin\theta_{\rm{UE}}\cos\omega_{\rm{UE}}\cos (\theta_{{\rm{PD,vert}},p}) \\
					C_1\sin\omega_{\rm{UE}}\cos\theta_{\rm{UE}} + C_2\cos\omega_{\rm{UE}}  + \sin\theta_{\rm{UE}}\sin\omega_{\rm{UE}} \cos (\theta_{{\rm{PD,vert}},p})\\
					-C_1\sin\theta_{\rm{UE}} + \cos\theta_{\rm{UE}} \cos (\theta_{{\rm{PD,vert}},p})\\
			\end{array} } \right]
		\end{aligned}
		\label{eq_normalV_randomOrien},
		$}
\end{equation}
where $C_1=\sin (\theta_{{\rm{PD,vert}},p}) \cos (\omega_{{\rm{PD,vert}},p})$ and $C_2=\sin (\theta_{{\rm{PD,vert}},p})\sin (\omega_{{\rm{PD,vert}},p})$.
Based on \eqref{eq_normalV_randomOrien}, after the random rotation,  the elevation angle of the $p$-th PD can be obtained as:
\begin{equation}
		\resizebox{0.85\hsize}{!}{$
			\theta_{{\rm{PD}},p}=\cos^{-1}\Big(-C_1\sin\theta_{\rm{UE}} + \cos\theta_{\rm{UE}} \cos \theta_{{\rm{PD,vert}},p}\Big),
		$}
\end{equation}
and the azimuth angle of the $p$-th PD can be expressed as \eqref{eq_omege_pd_p}. Therefore, the incidence angle of the $p$-th PD, $\psi_p$, can be obtained based on $\theta_{{\rm{PD}},p}$ and $\omega_{{\rm{PD}},p}$ as $\psi_p=\cos^{-1}(\frac{\mathbf{n}_{\rm{PD}}\cdot \mathbf{d}}{\lVert \mathbf{d} \rVert})$.

\begin{figure*}[!t]
	\normalsize
	\setcounter{mytempeqncnt}{\value{equation}}
	\setcounter{equation}{24}
	\begin{equation}
		\begin{split}
			p_{\rm{v}}(\Psi_{\rm{c}})&=\int\limits_{x_{\rm{UE}}}\int\limits_{y_{\rm{UE}}}\int\limits_{\theta_{\rm{UE}}}\int\limits_{\omega_{\rm{UE}}} V(x_{\rm{UE}},y_{\rm{UE}},\theta_{\rm{UE}},\omega_{\rm{UE}},\Psi_{\rm{c}})\frac{1}{X_{\rm{UE}}}\frac{1}{Y_{\rm{UE}}}\frac{1}{\Omega_{\rm{UE}}}f_{\theta}(\theta_{\rm{UE}})dx_{\rm{UE}}dy_{\rm{UE}}d\theta_{\rm{UE}}d\omega_{\rm{UE}}\\
			&=\int\limits_{x_{\rm{UE}}}\int\limits_{y_{\rm{UE}}}\int\limits_{\theta_{\rm{UE}}}\int\limits_{\omega_{\rm{UE}}} \frac{V(x_{\rm{UE}},y_{\rm{UE}},\theta_{\rm{UE}},\omega_{\rm{UE}},\Psi_{\rm{c}})}{X_{\rm{UE}}Y_{\rm{UE}}\Omega_{\rm{UE}}}f_{\theta}(\theta_{\rm{UE}})dx_{\rm{UE}}dy_{\rm{UE}}d\theta_{\rm{UE}}d\omega_{\rm{UE}}
		\end{split}
		\label{eq_Pv}
	\end{equation}
	\hrulefill
	\vspace*{4pt}
\end{figure*}

\subsection{Receiver Bandwidth vs PD Area}
The bandwidth of a PD is affected by its physical area, $A_{\rm{p}}$, and the PD thickness, $L_{\rm{p}}$. The capacitance of the each PD is denoted as $C_{\rm{r}}=\varepsilon_0 \varepsilon_{\rm{r}} \frac{A_{\rm{p}}}{L_{\rm{p}}}$, where $\varepsilon_0$ and $\varepsilon_{\rm{r}}$ are the permittivity of vacuum  and and the relative permittivity of silicon, respectively. 
The load resistance is defined as $R_{\rm{load}}$ while the hole velocity is denoted as $v_{\rm{p}}$. Therefore, the receiver bandwidth can be written as \cite{ReceiverDesign_1997}:
\begin{equation}
	\setcounter{equation}{22}
	B_{\rm{r}}=\frac{1}{\sqrt{(2\pi R_{\rm{load}}C_{\rm{r}})^2+\Big(\frac{L_{\rm{p}}}{0.443v_{\rm{p}}}\Big)^2}}.
	\label{eq_br}
\end{equation}
By solving $\frac{\partial B_{\rm{r}}}{\partial L_{\rm{p}}}=0$, the optimum $L_{\rm{p}}$ can be denoted as $L_{\rm{p,opt}}=\sqrt{0.886\pi R_{\rm{load}}\varepsilon_0 \varepsilon_{\rm{r}}A_{\rm{p}}v_{\rm{p}}}$.

\subsection{Visibility of an ADR}
The visibility of an ADR was first defined in \cite{ADR_ICC_ZENG}. An AP is visible to a PD
when the AP is within the FOV of the PD. Hence, at the location $\mathbf{p}_{\rm{UE}}$ and the orientation ($\theta_{\rm{UE}}, \omega_{\rm{UE}}$), the visibility factor between the $p$-th PD on the ADR and the $\alpha$-th AP  can be expressed as:
\begin{equation}
	\setcounter{equation}{23}
	\begin{split}
		&v_{\alpha,p}(x_{\rm{UE}},y_{\rm{UE}},\theta_{\rm{UE}},\omega_{\rm{UE}},\Psi_{\rm{c}})=
		\begin{cases}
			0, &  \psi_{\alpha,p}>\Psi_{\rm{c}}\\
			1, & \text{otherwise}
		\end{cases},\\
		&\text{and}~~\psi_{\alpha,p}=\arccos\Big(\frac{\mathbf{n}_{{\rm{PD}},p}(\theta_{\rm{UE}},\omega_{\rm{UE}})\cdot\mathbf{d}_{\alpha}}{\norm{\mathbf{d}_{\alpha}}}\Big),
	\end{split}
\end{equation}
where $\mathbf{d}_{\alpha}$= $(x_\alpha-x_{\rm{UE}},y_\alpha-y_{\rm{UE}},z_\alpha-z_{\rm{UE}})$ is the distance vector between the AP $\alpha$ and the UE. The dot product is denoted as $(\cdot)$ and $\norm{\cdot}$ is the norm operator.  In terms of ADR, an AP is visible to an ADR if and only if the AP is visible to at least one of the PDs on the ADR. Hence, for a given UE position and orientation, the visibility of the ADR can be written as \cite{ADR_ICC_ZENG}:
\begin{equation}
		\resizebox{0.85\hsize}{!}{$
	V(x_{\rm{UE}},y_{\rm{UE}},\theta_{\rm{UE}},\omega_{\rm{UE}},\Psi_{\rm{c}})=
	\begin{cases}
		1, & \text{if} \sum\limits_{\alpha\in\mathcal{A}}\sum\limits_{p\in\mathcal{P}}v_{\alpha,p}\neq 0 \\
		0, & \text{otherwise}
	\end{cases}.
	$}
\end{equation}
It is assumed that both $x_{\rm{UE}}$ and $ y_{\rm{UE}}$ follow a uniform distribution. The probability of visibility of an ADR is defined as the probability that there is at least one AP within the visible area of the ADR for all UE positions and orientations, and it can be expressed as \eqref{eq_Pv}, where $X_{\rm{UE}}$, $Y_{\rm{UE}}$ and $\Omega_{\rm{UE}}$ are the range of $x_{\rm{UE}}$, $y_{\rm{UE}}$ and $\omega_{\rm{UE}}$, respectively. Hence, it can be obtained that $X_{\rm{UE}}=\max(x_{\rm{UE}})-\min(x_{\rm{UE}})$, $Y_{\rm{UE}}=\max(y_{\rm{UE}})-\min(y_{\rm{UE}})$ and $\Omega_{\rm{UE}}=\max(\omega_{\rm{UE}})-\min(\omega_{\rm{UE}})$.

\section{The Optimum Field of View}
\label{section_Optimum_FOV}
\subsection{Optimization Problem}
In \eqref{eq_hlos}, the LOS channel gain $H_{\rm{LOS}}$ is a convex function of $\Psi_{\rm{c}}$ and decreases monotonically. Hence, the smaller the $\Psi_{\rm{c}}$, the higher the channel gain. However, when the $\Psi_{\rm{c}}$ of the PD is too small, there is a high chance that no APs are visible to the ADR and the LOS link cannot be constructed. Thus, there is a trade off between the LOS channel gain and visibility. The optimization problem is formulated as maximizing the LOS channel gain based on the constraint that the ADR should provide visibility for all UE locations. Thus, the optimization function is written as: 
\begin{equation}
	\setcounter{equation}{26}
	\begin{split}
		& \argmax_{\Psi_{\rm{c}}}~~ H_{\rm{LOS}}(\Psi_{\rm{c}}),\\
		&\text{subject to} ~~ p_{\rm{v}}(\Psi_{\rm{c}})=1.
	\end{split}
\end{equation}
The solution set of $p_{\rm{v}}(\Psi_{\rm{c}})=1$ is denoted as $\mathbb{R}$ and $\Psi_{\rm{c,min}}$ is the minimum value in $\mathbb{R}$. As $H_{\rm{LOS}}(\Psi_{\rm{c}})$ is a monotonically decreasing function, the maximum $H_{\rm{LOS}}(\Psi_{\rm{c}})$ is achieved when $\Psi_{\rm{c}}=\Psi_{\rm{c,min}}$. Hence, the optimization problem can be solved by finding the minimum value of $\Psi_{\rm{c}}$, $\Psi_{\rm{c,min}}$, which satisfies  $p_{\rm{v}}=1$. Based on \eqref{eq_Pv}, $\Psi_{\rm{c,min}}$ cannot be solved in a closed form. Therefore, in the following parts, we will study the ADRs' visible area on the ceilings to solve the solution set $\mathbb{R}$ and find a closed form for $\Psi_{\rm{c,min}}$.
\subsection{Coverage Area of ADR on the Ceiling}
\label{section_coverage}
The coverage area of a PR for a vertical-orientated UE is studied in \cite{ADR_ICC_ZENG}.

\begin{figure*}[!t]
	\normalsize
	\setcounter{mytempeqncnt}{\value{equation}}
	\setcounter{equation}{30}
	\begin{equation}
		\left[ {\begin{array}{c}
				x_{{\rm{ellipse}},p}\\
				y_{{\rm{ellipse}},p}\\ 
		\end{array} } \right]
		=
		\begin{cases}
			\left[ {\begin{array}{cc}
					\cos\omega_{{\rm{PD}},p} & -\sin\omega_{{\rm{PD}},p}\\
					\sin\omega_{{\rm{PD}},p} & \cos\omega_{{\rm{PD}},p}\\ 
			\end{array} } \right]
			\left[ {\begin{array}{c}
					x_{\rm{ellipse,1}}-x_{\rm{UE}}\\
					y_{\rm{ellipse,1}}-y_{\rm{UE}}\\ 
			\end{array} } \right] +
			\left[ {\begin{array}{c}
					x_{\rm{UE}}\\
					y_{\rm{UE}}\\ 
			\end{array} } \right] , & \text{if} \ 1 \leq p < N_{\rm{TPR}}\\\\
			\left[ {\begin{array}{c}
					x_{\rm{circle}}\\
					y_{\rm{circle}}\\ 
			\end{array} } \right] , & \text{if} \  p = N_{\rm{TPR}}
		\end{cases}
		\label{eq_ellipse_TPR}
	\end{equation}
	\setcounter{equation}{\value{mytempeqncnt}}
	\hrulefill
	\vspace*{4pt}
\end{figure*}

\begin{figure}[t]
	\centering
	\begin{subfigure}[ht]{\columnwidth}
		\centering
		\includegraphics[height=0.6\columnwidth]{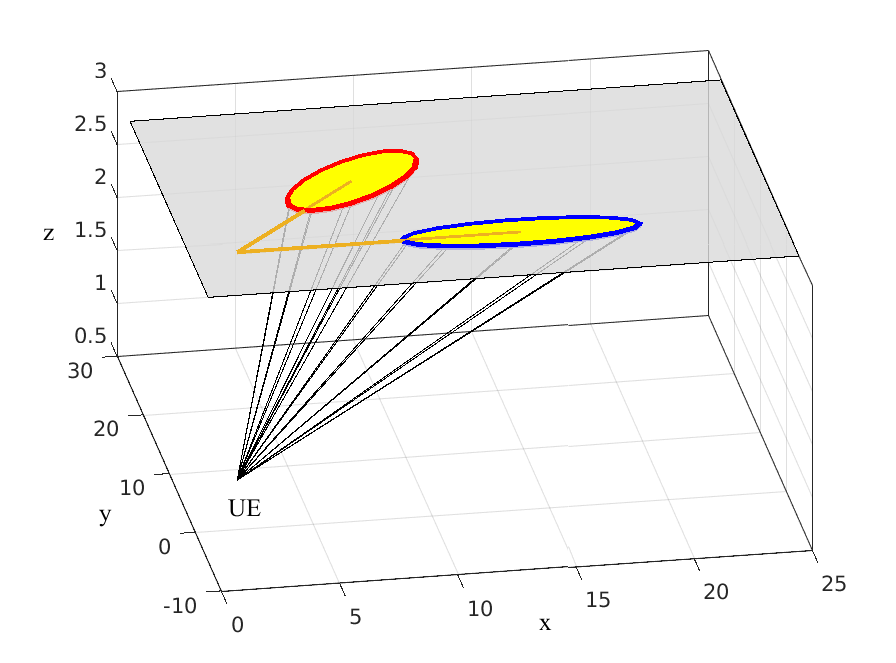}
		\caption{Visible area of PDs on the ceiling}
		\label{fig_ellipse3d}
	\end{subfigure}%
	\\
	\begin{subfigure}[ht]{\columnwidth}
		\centering
		\includegraphics[height=0.6\columnwidth]{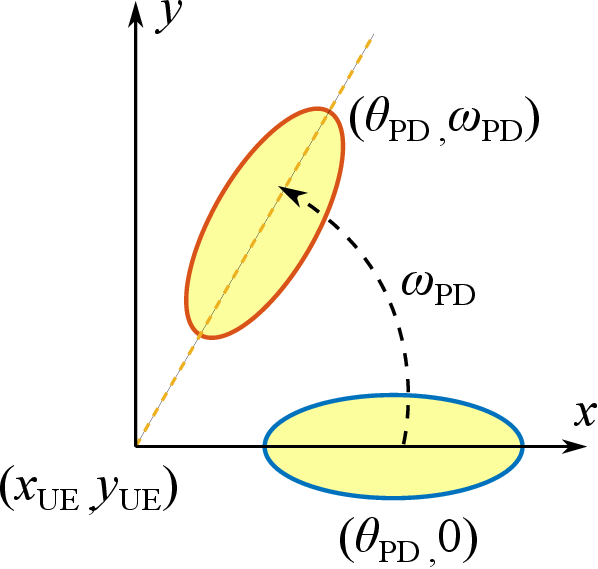}
		\caption{Representation of the visible area on the xy-plane.}
		\label{fig_ellipse}
	\end{subfigure}
	\caption{Visible area of PDs}
		\vspace{-10pt}
\end{figure}
Fig. \ref{fig_ellipse3d} demonstrates that the visible area of the PD mounted on the PR is an ellipse on the ceiling. Hence, the visible area of the 1-st PD, where $\theta_{\rm{PD,1}}=\Theta_{\rm{PD}}$ and $\omega_{\rm{PD,1}}=0$, is given by \cite{ADR_ICC_ZENG}:
\begin{equation}
	\frac{(x_{\rm{ellipse,1}}-x_{\rm{center}})^2}{a^2}+\frac{(y_{\rm{ellipse,1}}-y_{\rm{center}})^2}{b^2}=1,
	\label{eq_ellipse}
\end{equation}  
where 
\begin{equation}
	a=\frac{h\sin(2\Psi_{\rm{c}})}{\cos (2\Psi_{\rm{c}})+\cos (2\Theta_{\rm{PD}})},~~~
	b=\frac{\sqrt{2}h\sin(\Psi_{\rm{c}})}{\sqrt{\cos (2\Psi_{\rm{c}})+\cos (2\Theta_{\rm{PD}})}},
	\label{eq_ab}
\end{equation}

and 
\begin{equation}
	\resizebox{0.85\hsize}{!}{$
			x_{\rm{center}}=x_{\rm{UE}}+\frac{h\sin(2\Theta_{\rm{PD}})}{\cos (2\Psi_{\rm{c}})+\cos (2\Theta_{\rm{PD}})},~~~ y_{\rm{center}}=y_{\rm{UE}},
	$}
\end{equation}
where $h$ is the vertical distance between the AP and UE. The detailed proof is given in Appendix \ref{appendix1}. Fig. \ref{fig_ellipse} depicts that the shape of the visible area of the $p$-th PD can be obtained by rotating the 1-st PD around $(x_{\rm{UE}},y_{\rm{UE}})$ with an angle of $\omega_{{\rm{PD}},p}$, which can be represented as:
\begin{equation}
	\begin{split}
		&\left[ {\begin{array}{c}
				x_{{\rm{ellipse}},p}-x_{\rm{UE}}\\
				y_{{\rm{ellipse}},p}-y_{\rm{UE}}\\ 
		\end{array} } \right]
		=R_{xy}(\omega_{{\rm{PD}},p})\left[ {\begin{array}{c}
				x_{\rm{ellipse,1}}-x_{\rm{UE}}\\
				y_{\rm{ellipse,1}}-y_{\rm{UE}}\\ 
		\end{array} } \right]
		=\\
		&\left[ {\begin{array}{cc}
				\cos\omega_{{\rm{PD}},p} & -\sin\omega_{{\rm{PD}},p}\\
				\sin\omega_{{\rm{PD}},p} & \cos\omega_{{\rm{PD}},p}\\ 
		\end{array} } \right]
		\left[ {\begin{array}{c}
				x_{\rm{ellipse,1}}-x_{\rm{UE}}\\
				y_{\rm{ellipse,1}}-y_{\rm{UE}}\\ 
		\end{array} } \right].
		\label{eq_ellipse_PR}
	\end{split}	
\end{equation}
The TPR can be seen as the combination of a PR, where $N_{\rm{PR}}=N_{\rm{TPR}}-1$, and a central PD. When the UE is facing vertically upward, the visible area of the central PD is a circle. Therefore, the shape of the visible area of the $p$-th PD on a TPR is given by \eqref{eq_ellipse_TPR}, where $x_{\rm{circle}}^2 + y_{\rm{circle}}^2 = (h\tan \Psi_{\rm{c}})^2$.

\begin{figure}[!h]
	\centering
	\includegraphics[width=\columnwidth]{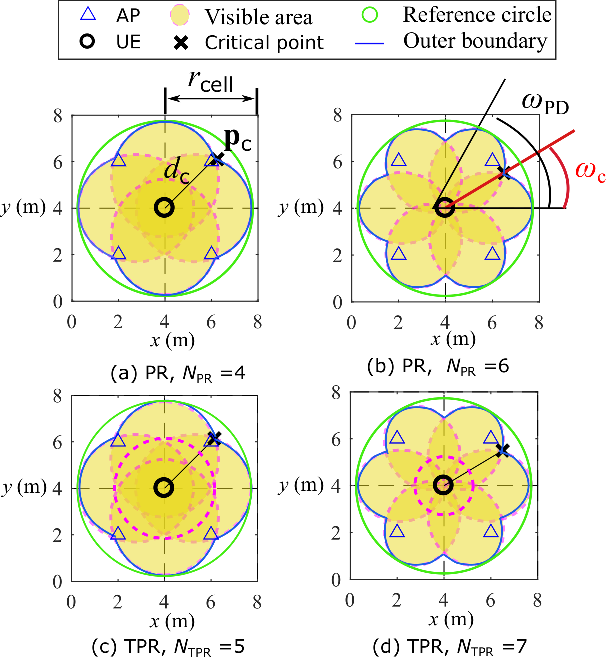}
	\caption{Coverage area of PRs and TPRs with different number of PDs.}
	\label{fig_coverage_all}
		\vspace{-10pt}
\end{figure}

\subsection{Lower Bound of FOV}
\label{section_lowerBoundFOV}
For a fixed UE location, the ADR has the smallest coverage area on the ceiling when vertically orientated. In other words, we will investigate the worst condition, i.e. the situation that an ADR is positioned vertically upward which provides the smallest coverage area on the ceiling. Under other orientation scenarios, the coverage area is larger. Based on \eqref{eq_ellipse_PR} and \eqref{eq_ellipse_TPR}, Fig. \ref{fig_coverage_all} illustrates the visible area of 4 different types of ADRs when the UE is at the cell corner, that is to say, the cross-point of four LiFi cells. The blue curve is the outer boundary of the visible area. On the outer boundary, the points that have the shortest distance to the UE are defined as critical points, $\mathbf{p}_{\rm{c}}$. $d_{\rm{c}}$ denotes the horizontal distance between $\mathbf{p}_{\rm{c}}$ and the UE. To ensure $p_{\rm{v}}=1$, there are two constraints and the detailed explanation of these constraints are given as follows.
\subsubsection{Constraint $\Rmnum{1}$}\textit{The central area above the ADR should be visible to the ADR.}
\begin{figure}[!ht]
	\centering
	\begin{subfigure}[ht]{\columnwidth}
		\centering
		\includegraphics[height=0.5\columnwidth]{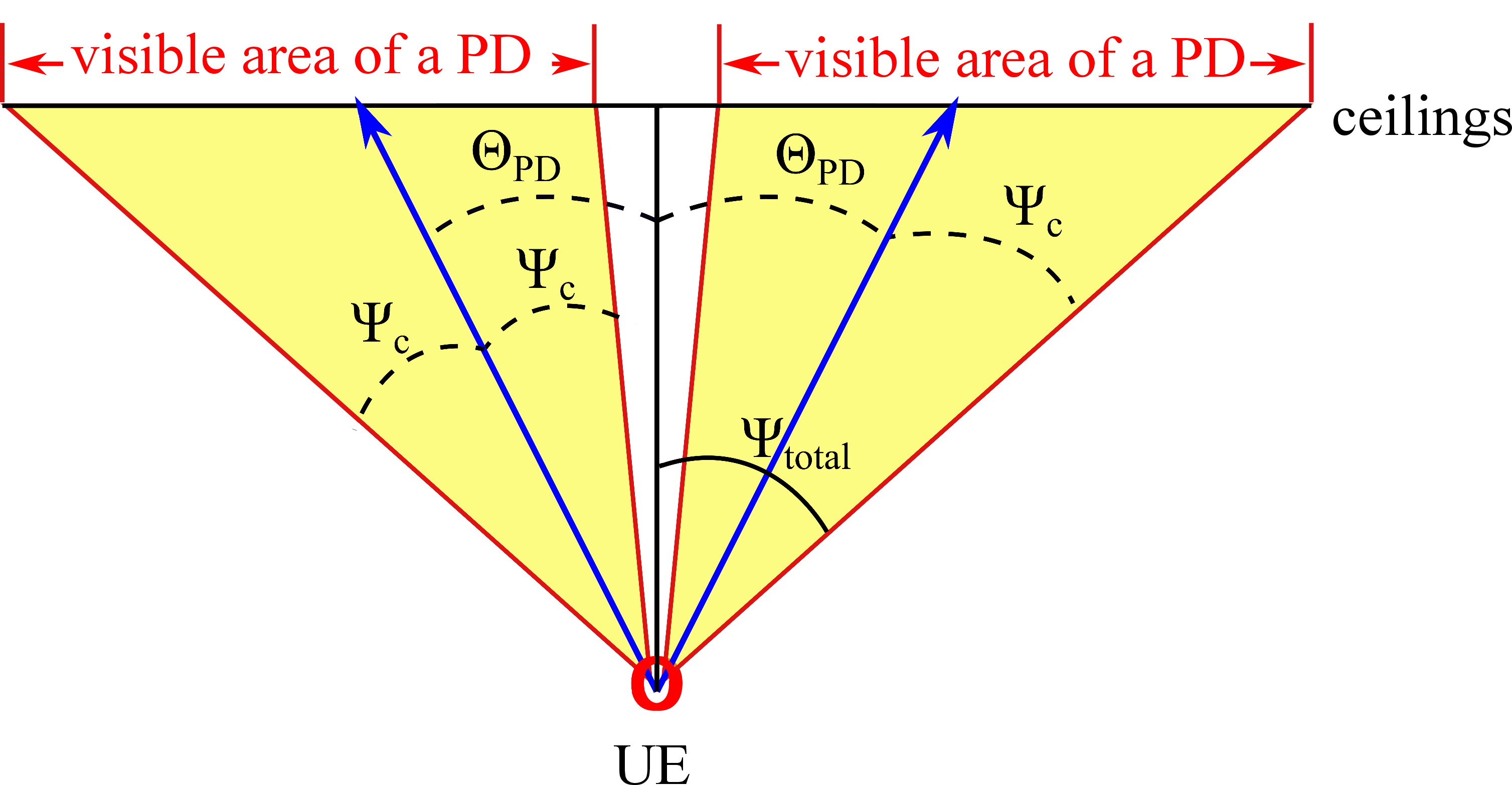}
		\caption{PR}
		\label{fig_coverage_xz}
	\end{subfigure}%
	\\
	\begin{subfigure}[ht]{\columnwidth}
		\centering
		\includegraphics[height=0.45\columnwidth]{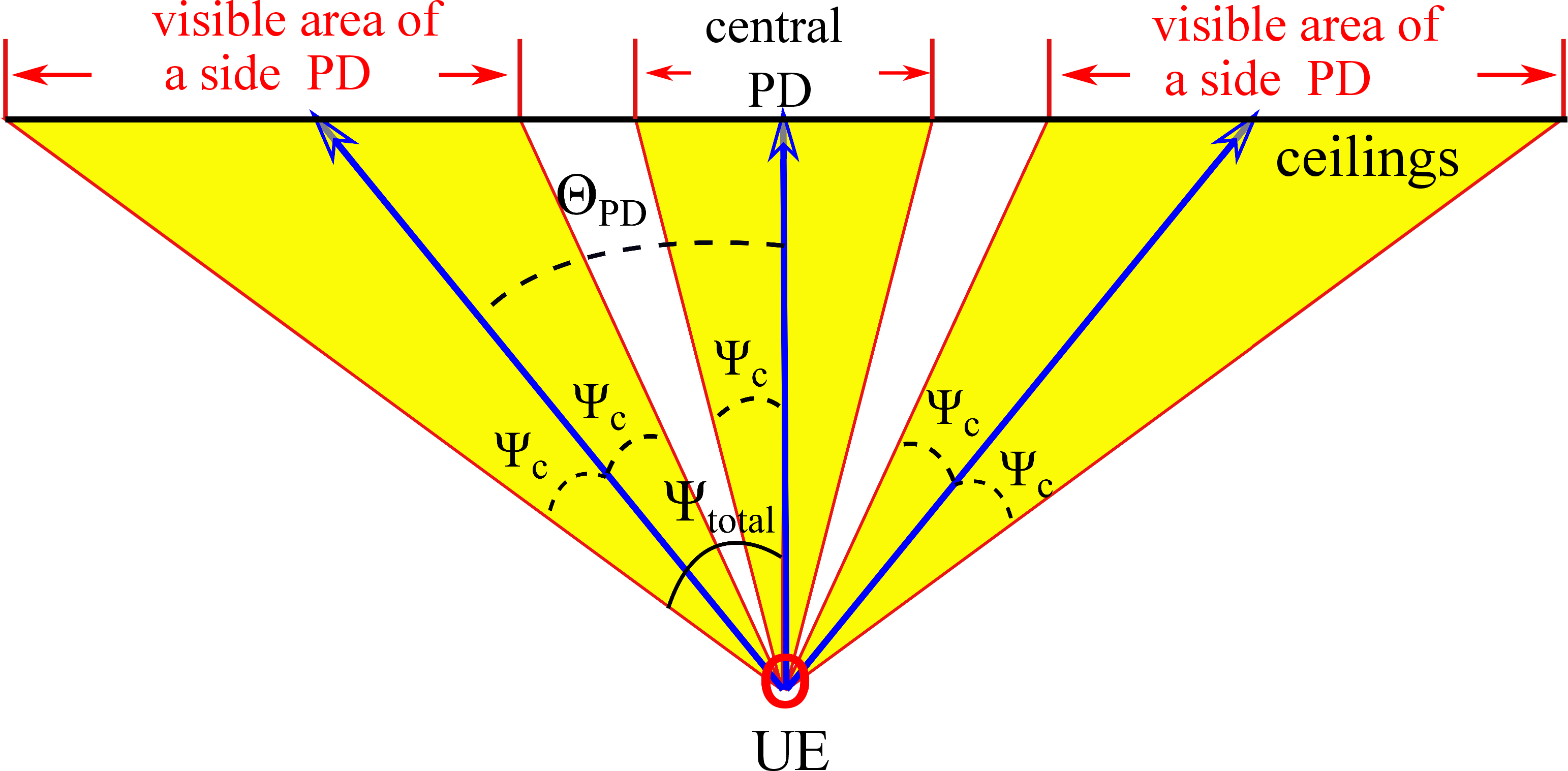}
		\caption{TPR}
		\label{fig_coverage_xz_TPR}
	\end{subfigure}
	\caption{Visible area of ADR in $xz$-plane.}
	\label{fig_SS_config}
	\vspace{-10pt}
\end{figure}
As shown in Fig. \ref{fig_coverage_xz}, the total FOV of an ADR is represented as $\Psi_{\rm{total}}$, which can be written as:
\begin{equation}
	\setcounter{equation}{32}
	\Psi_{\rm{total}}=\Psi_{\rm{c}}+\Theta_{\rm{PD}}.
	\label{eq_Psi_total}
\end{equation}
In terms of PRs, if $\Theta_{\rm{PD}} \geq \Psi_{\rm{c}}$, then the central part is not covered by the visible area of the ADR as manifested in Fig. \ref{fig_coverage_xz}. If the UE is in the cell center, then no APs will be visible to the ADR. Hence, the condition $\Theta_{\rm{PD}} \leq \Psi_{\rm{c}}$ should be satisfied so that the area directly above the UE is covered by the visible area of the ADR. Based on this constraint and \eqref{eq_Psi_total}, the lower bound of $\Psi_{\rm{c}}$ can be obtained as:
\begin{equation}
	\Psi_{\rm{c1,min}} = \frac{\Psi_{\rm{total}}}{2}.
	\label{eq_constraint1_PR}
\end{equation}

With respect to the TPR, the area directly above the UE is covered by the central PD orientating vertically upwards as illustrated in Fig. \ref{fig_coverage_xz_TPR}. The concern should be the central coverage gap between the central PD and the side PDs. Therefore, $\Theta_{\rm{PD}} \leq 2\Psi_{\rm{c}}$ is required to ensure there is no gap between them. By substituting this constraint into \eqref{eq_Psi_total}, it can be derived that:
\begin{equation}
	\Psi_{\rm{c1,min}} = \frac{\Psi_{\rm{total}}}{3}.
	\label{eq_constraint1_TPR}
\end{equation}

\begin{figure}[!t]
	\centering
	\includegraphics[width=0.6\columnwidth]{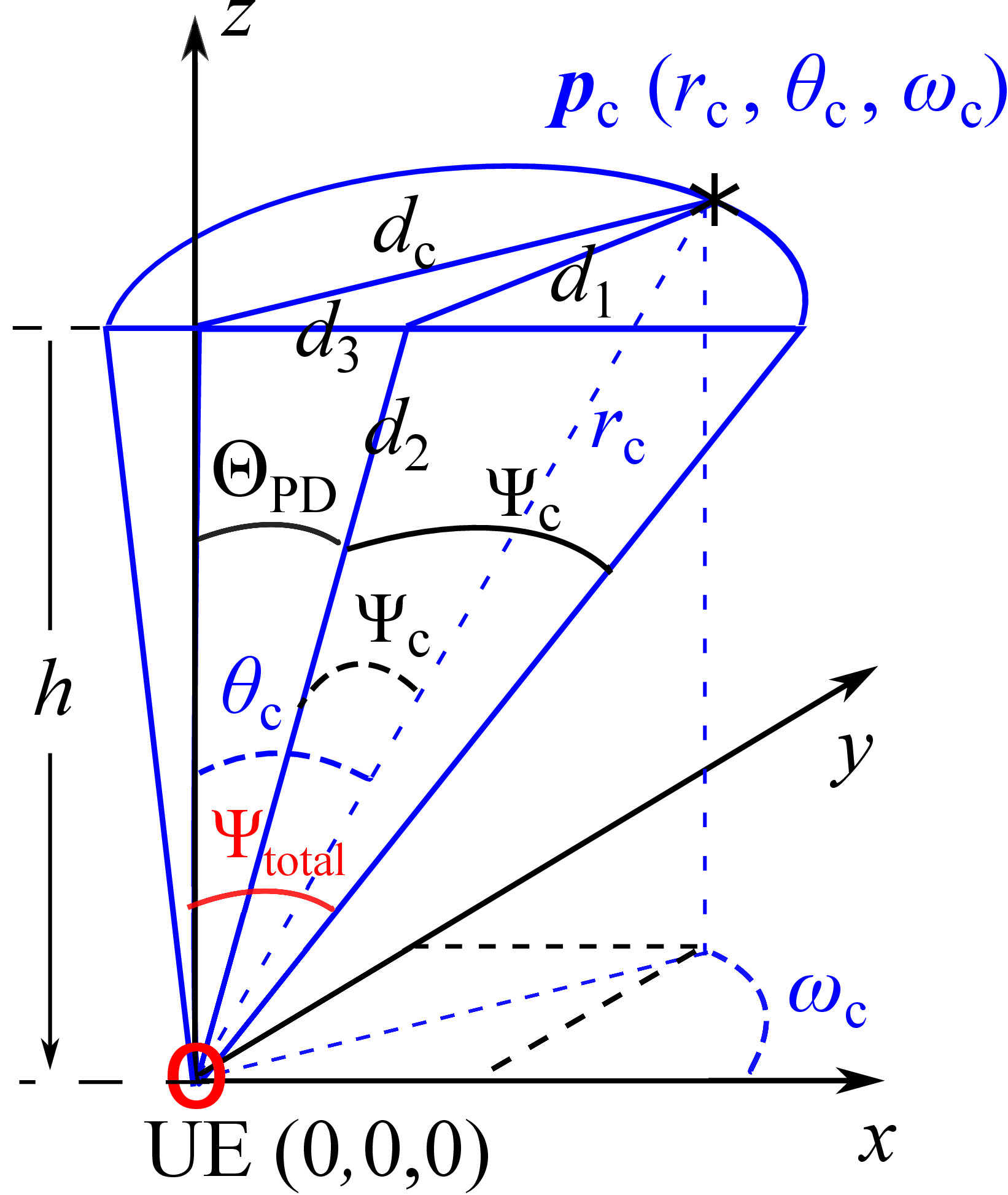}
	\caption{ The geometrical representation of $\Psi_{\rm{c}}$, $\Theta_{\rm{PD}}$, $\Psi_{\rm{total}}$ and $d_{\rm{c}}$ in the spherical coordinate system.}
	\label{fig_3d_coverage}
		\vspace{-10pt}
\end{figure}

\subsubsection{Constraint $\Rmnum{2}$} \textit{The outer boundary of the visible area should be large enough.}
The side length of a square cell is denoted as $r_{\rm{cell}}$ as shown in \mbox{Fig. \ref{fig_coverage_all}}. The horizontal distance between the UE and the $\alpha$-th AP is denoted as $d_{{\rm{h}},\alpha}$. When the UE is at the cell corner, $d_{{\rm{h}},\alpha}$ = $\frac{\sqrt{2}}{2}r_{\rm{cell}}$ for any $\alpha \in \mathcal{A}$. With the decrease of $\Psi_{\rm{c}}$, the outer boundary of the visible area will decrease, which means $d_{\rm{c}}$ will decrease.  If $d_{\rm{c}}$ is smaller than the horizontal distance from the AP to the cell corner, which is $\frac{\sqrt{2}}{2}r_{\rm{cell}}$, there will be no APs within the visible area of the ADR for cell-corner users. Therefore, to ensure that at least one AP is visible to the cell-corner UE, $d_{\rm{c}}$ should be larger than the horizontal distance from the AP to the cell corner. By moving towards any direction, due to the symmetry, the cell-corner UE will get closer to at least one AP. In other words, 
\begin{equation}
		\resizebox{\hsize}{!}{$
	d_{\rm{c,min}}=\max_{x_{\rm{UE}},y_{\rm{UE}}}\Big( \min_{\alpha}\big(d_{{\rm{h}},\alpha}\big)\Big) =\frac{\sqrt{2}}{2}r_{\rm{cell}},~~\text{subject to} ~~  \alpha\in\mathcal{A}
	\label{eq_dc_min_single}.
	$}
\end{equation}
That is to say, if there is at least one AP inside the outer boundary of the visible area for the cell-corner UE, then, when the UE moves to other locations, the AP will still be inside the outer boundary of the visible area. Hence, to meet the condition $p_{\rm{v}}=1$, it is required that $d_{\rm{c}} \geq \frac{\sqrt{2}}{2}r_{\rm{cell}} $. Also, it can be seen from Fig. \ref{fig_coverage_all} that $\mathbf{p}_{\rm{c}}$ is always inside the green reference circle, which has a radius of $h\tan(\Psi_{\rm{total}})$.  Hence, $ d_{\rm{c,min}} \leq d_{\rm{c}} \leq h\tan(\Psi_{\rm{total}})$, where $d_{\rm{c,min}}=\frac{\sqrt{2}}{2}r_{\rm{cell}}$.

Fig. \ref{fig_3d_coverage} presents the geometrical relationship in a spherical coordinate system. The coordinate of $\mathbf{p}_{\rm{c}}$ is represented as $(r_{\rm{c}}, \theta_{\rm{c}}, \omega_{\rm{c}})$ 
and $\omega_{\rm{c}}=\frac{\omega_{{\rm{PD}},p}}{2} |_{p=2}$ for both PRs and TPRs. The geometrical relationships among $d_1, d_2$ and $d_3$, illustrated in Fig. \ref{fig_3d_coverage}, can be represented as:
\begin{equation}
	d^2_1=d^2_2+r^2_{\rm{c}}-2d_2r_{\rm{c}}\cos(\Psi_{\rm{c}}),
	\label{eq_d12}
\end{equation}
\begin{equation}
	d^2_1=d^2_3+d^2_{\rm{c}}-2d_3d_{\rm{c}}\cos(\omega_{\rm{c}}),
	\label{eq_d13}
\end{equation}
\begin{equation}
	d^2_2=h^2+d^2_3,~~
	r^2_{\rm{c}}=h^2+d^2_{\rm{c}},~~
	d_3=h\tan(\Theta_{\rm{PD}}).
	\label{eq_d2_rc_d3}
\end{equation}
According to \eqref{eq_d13}, \eqref{eq_d12} and \eqref{eq_d2_rc_d3}, the lower bound of $\Psi_{\rm{c}}$ set by Constraint $\rm{\Rmnum{2}}$ is derived in Appendix \ref{appendix2} and is represented as:
\begin{equation}
	\Psi_{\rm{c2,min}}= 
	\begin{dcases}
		F_2(d_{\rm{c2}}),& \text{if} \ d_{\rm{c,min}} \leq d_{\rm{c2}} \\
		F_2(d_{\rm{c,min}}),              & \text{otherwise}
	\end{dcases},
	\label{eq_Psi_c2_min}
\end{equation}
where 
\begin{equation}
		\resizebox{0.85\hsize}{!}{$
			F_2(d_{\rm{c}})=\Psi_{\rm{total}}-\tan^{-1}\frac{\sqrt{h^2+d^2_{\rm{c}}}\cos(\Psi_{\rm{total}})-h}{{d_{\rm{c}}\cos(\omega_{\rm{c}})-\sqrt{h^2+d^2_{\rm{c}}}\sin(\Psi_{\rm{total}})}},
		$}
	\label{eq_lb_F2}
\end{equation}
and 
\begin{equation}
	d_{\rm{c2}}=\frac{h\cos(\omega_{\rm{c}})\sin(\Psi_{\rm{total}})}{\cos(\Psi_{\rm{total}})+\sin(\omega_{\rm{c}})}.
\end{equation}

\begin{figure*}[!t]
	\normalsize
	\setcounter{mytempeqncnt}{\value{equation}}
	\setcounter{equation}{45}
	\begin{equation}
		d_{\rm{c,min}}=\max_{x_{\rm{UE}},y_{\rm{UE}}} \Big( \min_{\alpha}\big(d_{{\rm{h}},\alpha}\big)\Big) =\\
		\begin{dcases}
			\sqrt{\big(\frac{r_{\rm{cell}}}{2}-\frac{d_{\rm{source}}}{2}\big)^2+\big(\frac{r_{\rm{cell}}}{2}\big)^2},& \text{if} \ d_{\rm{source}} \leq \frac{r_{\rm{cell}}}{2} \\
			\sqrt{\big(\frac{r_{\rm{cell}}}{2}\big)^2+\big(\frac{d_{\rm{source}}}{2}\big)^2},     & \text{otherwise}
		\end{dcases},~~~~\text{subject to} ~~  \alpha\in\mathcal{A}.
		\label{eq_dc_min_double}
	\end{equation} 
	\setcounter{equation}{\value{mytempeqncnt}}
	\hrulefill
	\vspace*{4pt}
\end{figure*}

\begin{figure}[!t]
	\centering
	\begin{subfigure}[h]{0.45\textwidth}
		\centering
		\includegraphics[width=0.8\columnwidth]{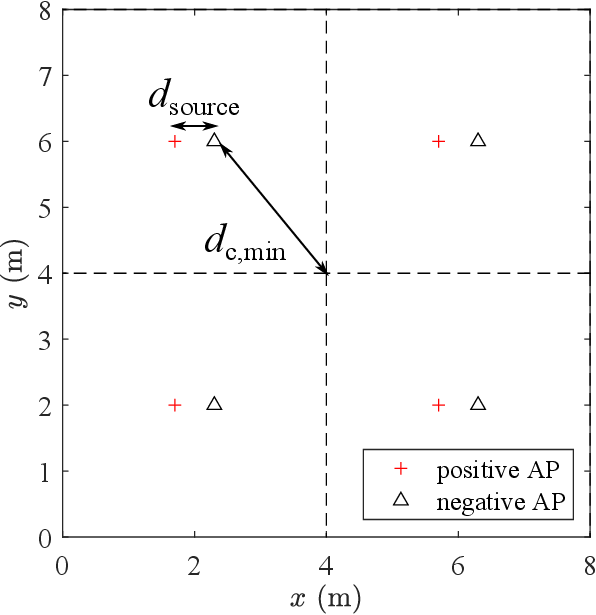}
		\caption{$d_{\rm{source}}\leq \frac{r_{\rm{cell}}}{2}$}
		\label{fig_2source_network_small}
	\end{subfigure}%
	\\
	\begin{subfigure}[h]{0.45\textwidth}
		\centering
		\includegraphics[width=0.8\columnwidth]{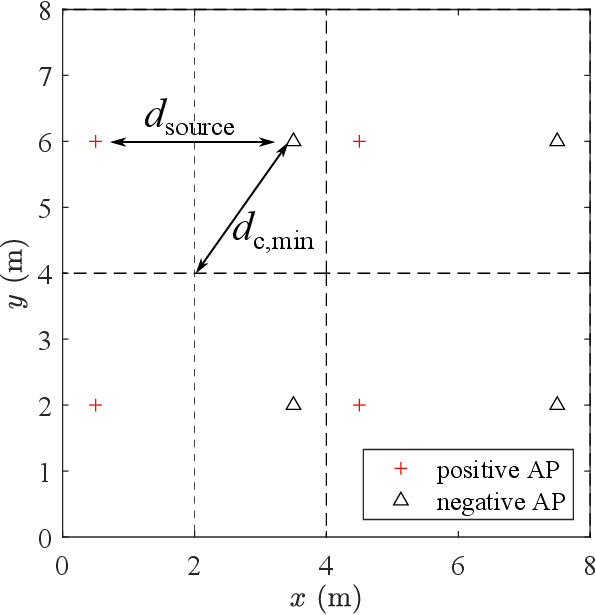}
		\caption{$d_{\rm{source}}> \frac{r_{\rm{cell}}}{2}$}
		\label{fig_2source_network_large}
	\end{subfigure}
	\caption{Double source cell configuration.}
	\label{fig_2source_network}
	\vspace{-10pt}
\end{figure} 

\subsubsection{Summary}
Based on \eqref{eq_constraint1_PR}, \eqref{eq_constraint1_TPR} and \eqref{eq_Psi_c2_min}, the lower bound of $\Psi_{\rm{c}}$ can be expressed as:
\begin{equation}
	\Psi_{\rm{c,min}}= \text{max}(\Psi_{\rm{c1,min}},\Psi_{\rm{c2,min}}).
	\label{eq_lb_PR}
\end{equation}
Therefore, the solution set $\mathbb{R}$ is $\Psi_{\rm{c,min}}\leq\Psi_{\rm{c}}\leq\Psi_{\rm{total}}$. For different numbers of PDs on the PR, the optimum FOV is $\Psi_{\rm{c,min}}$ as the  FOV gets smaller, the higher the channel gain and received signal power.


\section{Double Source Cell Configuration}
\label{section_DS_FS}
In the conventional SS cell configuration, each cell is equipped with a single AP in the cell center. The double source (DS) cell configuration is proposed to further exploit the spatial diversity of the ADR in \cite{DoubSource_GlobCom_Zhe}. As demonstrated in Fig. \ref{fig_2source_network}, each LiFi AP consists of two sources which transmit the same information signals but with opposite polarity. These two sources are termed as the positive source and the negative source, which transmit the time domain signal $s_{\rm{pos}}(t)$ and $s_{\rm{neg}}(t)$ respectively. In a single optical cell, the received optical signal at a PD is denoted as \cite{DoubSource_GlobCom_Zhe}:
\begin{equation}
	s_{\rm{sum}}(t)=s_{\rm{pos}}(t)H_{\rm{pos}}+s_{\rm{neg}}(t)H_{\rm{neg}},
\end{equation}
where $H_{\rm{pos}}$ is the channel gain between the positive source and the PD; $H_{\rm{neg}}$ is the channel gain between the negative source and the PD. For a fair comparison, the total transmitting power for the SS system and DS system should be the same. Hence, the transmit power of each source is halved when the DS configuration is applied and the received optical power at the PD can be written as \cite{DoubSource_GlobCom_Zhe}:
\begin{equation}
	P_{\rm{rx}}=\frac{P_{\rm{tx}}}{2}|H_{\rm{pos}}-H_{\rm{neg}}|=\frac{P_{\rm{tx}}\Delta H}{2}. 
\end{equation}

Generally, the receiver is closer to the desired AP than the interfering AP. For the desired AP, due to the narrow FOV of ADRs, one PD can hardly receive LOS signals from both the positive source and negative source simultaneously, and only one appears as the LOS channel gain. In respect of the interfering AP, the channel gains $H_{\rm{pos}}$ and $H_{\rm{neg}}$ are both NLOS. Hence, the difference between $H_{\rm{pos}}$ and $H_{\rm{neg}}$ is small and the interference is attenuated accordingly. Therefore, the double source cell configuration can suppress the signal power from interfering APs \cite{DoubSource_GlobCom_Zhe}.  As the LOS interference can be mitigated by the narrow FOV of the ADR and the NLOS interference can be mitigated due to the adoption of the DS configuration, the SINR of user $\mu$ on subcarrier $m$ can be approximated by: 
\begin{equation}
	\centering
	\tilde{\gamma}_{\mu,m}\approx\frac{(\sum\limits_{p=1}^{N_{\rm{PD}}} w_p \tau \frac{P_{\rm{tx}}}{2} \Delta H_{\alpha_{\rm{s}},\mu,p})^2}{\sum\limits_{p=1}^{N_{\rm{PD}}}{w_p}^2\kappa^2N_0B_{\rm{L}}(M-2)/M},
	\label{eq_appro_SINR}
\end{equation}
where $\Delta H_{\alpha_{\rm{s}},\mu,p}$ is the overall DC channel gain between the PD $p$ of user $\mu$ and the serving AP $a_{\rm{s}}$ in the DS system.
As manifested in Fig. \ref{fig_2source_network}, $d_{\rm{c,min}}$ will vary according to the distance between the positive and negative sources, which can be represented as \eqref{eq_dc_min_double}. Therefore, the lower bound of $\Psi_{\rm{c}}$ for the double source cell system can be calculated based on \eqref{eq_Psi_c2_min} - \eqref{eq_lb_PR}.

\begin{figure*}[!t]
	\centering
	\begin{subfigure}[h]{0.45\textwidth}
		\centering
		\includegraphics[width=\columnwidth]{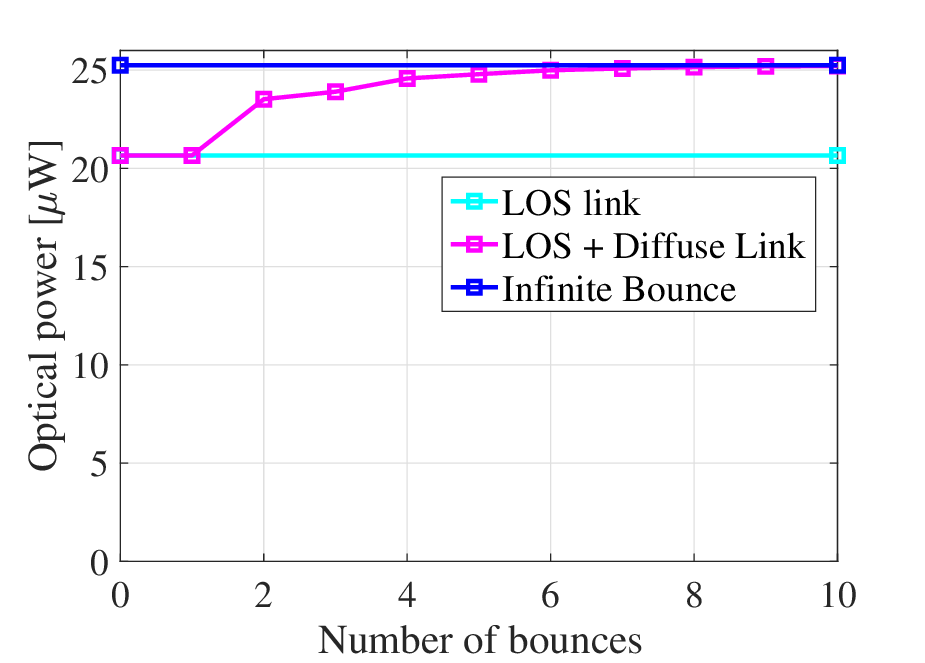}
		\caption{$x_{\rm{UE}}=6, \ y_{\rm{UE}}=6$, \ $\theta=0^\circ$, \ $\omega=0^\circ$}
		\label{optical_power_comparison_a}
	\end{subfigure}%
	~
	\begin{subfigure}[h]{0.45\textwidth}
		\centering
		\includegraphics[width=\columnwidth]{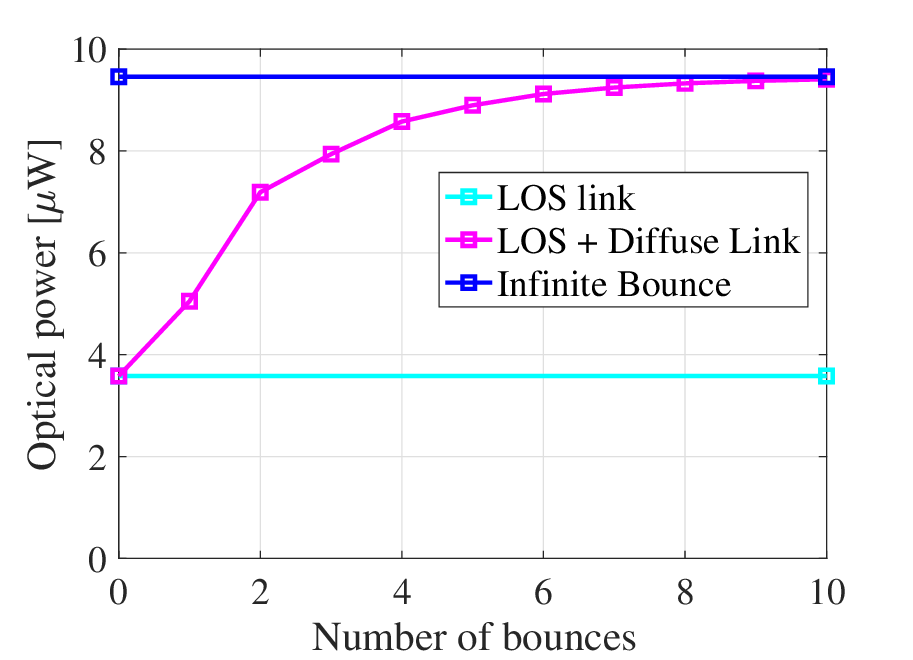}
		\caption{$x_{\rm{UE}}=6, \ y_{\rm{UE}}=6$, \ $\theta=60^\circ$, \ $\omega=180^\circ$}
		\label{optical_power_comparison_b}	
	\end{subfigure}\\
	\begin{subfigure}[h]{0.45\textwidth}
		\centering
		\includegraphics[width=\columnwidth]{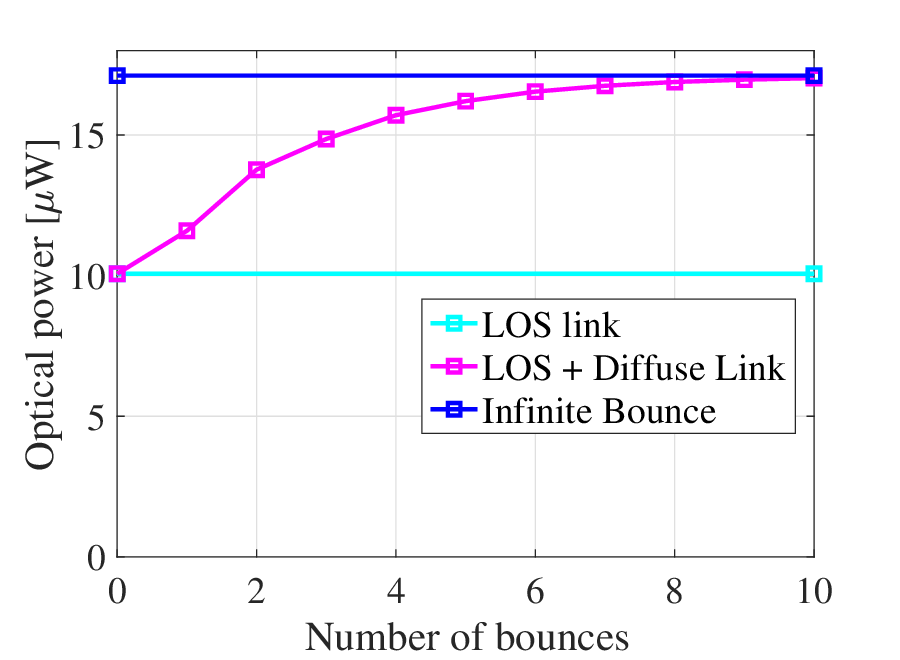}
		\caption{$x_{\rm{UE}}=1, \ y_{\rm{UE}}=1$, \ $\theta=0^\circ$, \ $\omega=0^\circ$}
		\label{optical_power_comparison_c}
	\end{subfigure}%
	~ 
	\begin{subfigure}[h]{0.45\textwidth}
		\centering
		\includegraphics[width=\columnwidth]{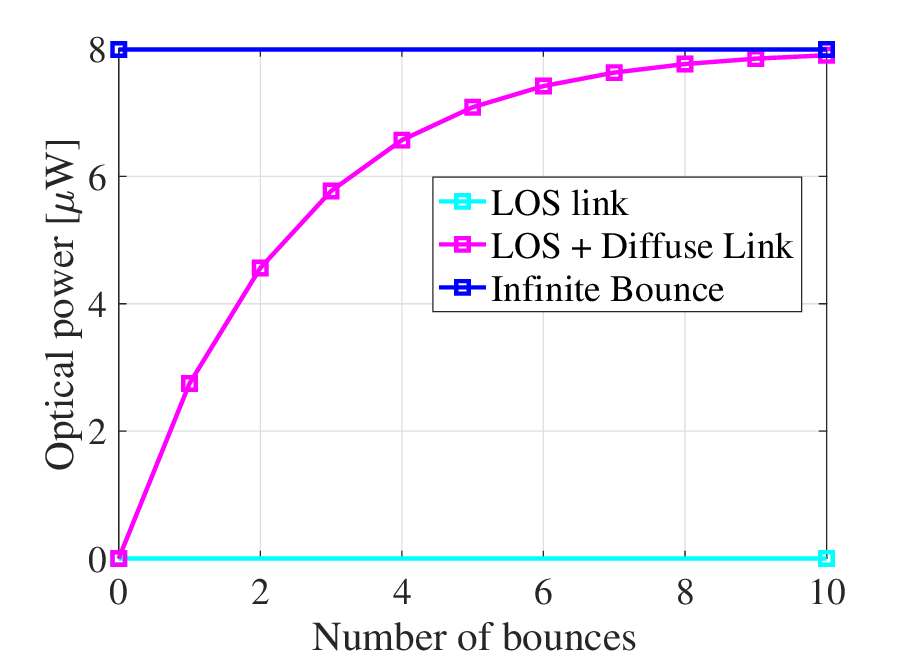}
		\caption{$x_{\rm{UE}}=1, \ y_{\rm{UE}}=1$, \ $\theta=45^\circ$, \ $\omega=180^\circ$}
		\label{optical_power_comparison_d}
	\end{subfigure}
	\caption{Comparisons among received optical power for LOS link, LOS + diffuse link up to order $L$ and infinite reflections in 2 different positions with different orientations.}
	\label{fig_optical_power_comparison}
	\vspace{-10pt}
\end{figure*}

\section{Results and Discussions}
\label{section_results}
\subsection{Simulation Setups}
As shown in Fig. \ref{fig_coverage_all}, an 8 m $\times$ 8 m $\times$ 3 m indoor office scenario is considered in this study, where 4 LiFi APs are deployed following a square topology. All of the users are uniformly distributed in the room and move randomly following the random waypoint model \cite{MDSorientation}. To make a fair comparison, the total physical area, $A_{\rm{t}}=N_{\rm{PD}}A_{\rm{p}}$, of the ADRs  should be the same. Hence, the physical area $A_{\rm{p}}$ on each PD decreases when the number of PDs increases. The other parameters used in the simulations are listed in Table \ref{table2}.

\begin{table}[!h]
	\centering
	\caption{Parameters Lists}
	\resizebox{\hsize}{!}{$
		\begin{tabular}{|l|l|l|}
			\hline
			Parameter &  Symbol  & Value \\
			\hline
			\hline
			Transmitted optical power per AP & $P_{\rm{tx}}$ & 10 W\\
			\hline
			Modulated bandwidth for LED & $B$ & 20 MHz\\
			\hline
			Physical are of the single PD receiver& $A_{\rm{p}}$ &  1 $\rm{cm^2}$\\
			\hline
			FOV of the single PD receiver & $\Psi_{\rm{c}}$ &  $60^{\circ}$\\				
			\hline
			The total FOV of an ADR  & $\Psi_{\rm{total}}$ & $60^\circ$\\
			\hline
			Half-intensity radiation angle & $\Phi_{1/2}$ &  $60^{\circ}$\\
			\hline
			PD responsivity & $\tau$ & 0.5 A/W\\
			\hline
			Noise power spectral density  & $N_0$ & $10^{-21}$ A$^2$/Hz\\
			\hline
			Vertical distance between APs and UEs & $h$ & $2.15$ m\\  
			\hline
			Wall reflectivity & $\rho_{\rm{wall}}$ & 0.8\\  
			\hline
			Ceiling reflectivity & $\rho_{\rm{ceiling}}$ & 0.8\\  
			\hline
			Floor reflectivity & $\rho_{\rm{floor}}$ & 0.3\\  
			\hline
			Refractive index &  $n_{\rm{ref}}$ & 1.5 \\
			\hline
			Optical filter gain & $G$ & 1\\
			\hline
			Permittivity of vacuum & $\varepsilon_0$ & $8.854 \times 10^{-12}$ F$\cdot {\rm{m}}^{-1}$\\
			\hline
			Relative permittivity of silicon & $\varepsilon_{\rm{r}}$ & 11.68\\
			\hline
			Hole velocity & $v_{\rm{p}}$ & $4.8 \times 10^{4}$ m/s\\
			\hline
		\end{tabular}
		$}
	\label{table2}
	\vspace{-10pt}
\end{table}

\subsection{Importance of Reflection and Orientation}

Fig. \ref{fig_optical_power_comparison} shows the received optical power in two different locations with different orientations. The room setup and the SS deployment is as shown in Fig. \ref{fig_SS_config} and a single PD receiver with a FOV of $60^\circ$ is used.  The received optical power from the LOS and NLOS link can be calculated based on \eqref{eq_hlos} and \eqref{eq_total_Transfer_func}, respectively. The ratio of the received optical power from the LOS signal link to the total received optical power is represented as $p_{\rm{los}}$. In Fig. \ref{optical_power_comparison_a} and \ref{optical_power_comparison_b}, the $p_{\rm{los}}$ of cell-center UEs at (6,6) degrades substantially when the orientation changes. In Fig. \ref{optical_power_comparison_c}, the optical power from the diffuse link occupies more than $40\%$ of the total optical power when $x_{\rm{UE}}=1,y_{\rm{UE}}=1$. Nevertheless, in Fig. \ref{optical_power_comparison_d}, by changing the device orientation, there is no LOS signal and only the signal from the diffuse links can be received. Therefore, the device orientation has a great impact on the received single power and thus cannot be ignored. In addition, both the LOS link and diffuse link should be considered to analyze the performance of a multi-cell visible light communication system. Fig. \ref{fig_optical_power_comparison} demonstrates that when the number of bounces is more than 5, the corresponding paths make a minor contribution to the total optical power. Hence, to reduce the computational complexity while maintaining high channel estimation accuracy, a light reflection order of $L=5$ and the orientation model proposed in \cite{MDSorientation} are considered for the following simulations. 
\begin{figure}[!t]
	\centering
	\includegraphics[width=\columnwidth]{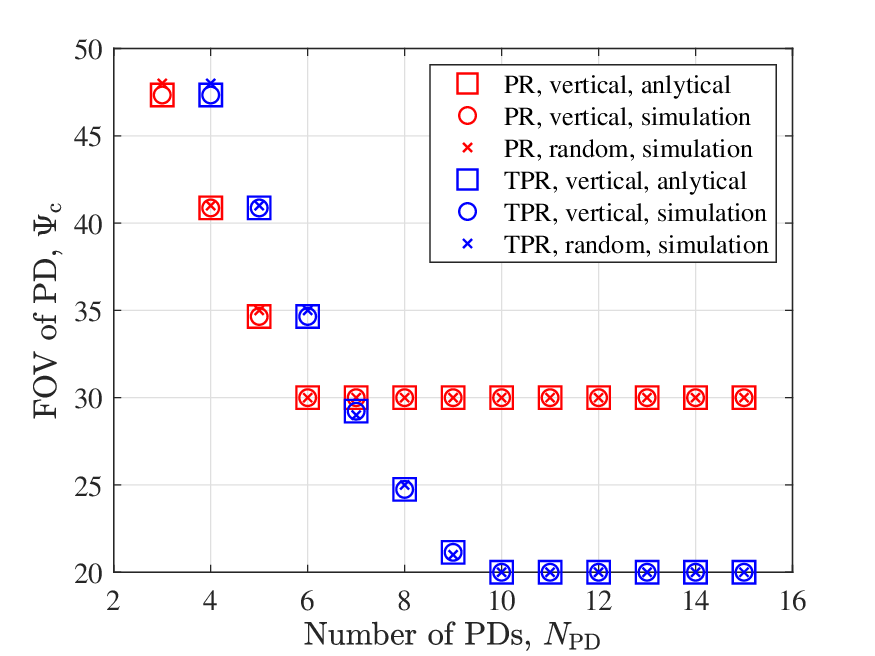}
	\caption{ The relationship between $\Psi_{\rm{c,min}}$ and the number of PDs, $N_{\rm{PD}}$, on an ADR for the SS system.}
	\label{fig_FOV_vs_N}
	\vspace{-10pt}
\end{figure}
\subsection{Performance Analysis for SS cells}
1) \textit{Lower bound of FOV}: Fig. \ref{fig_FOV_vs_N} manifests the relationship between the lower bound of the FOV, $\Psi_{\rm{c,min}}$, and the number of PDs. The analytical results are calculated based on the lower bound given in \eqref{eq_lb_PR}. 
The Monte-Carlo simulations for UEs with a vertical orientation can be carried out based on \eqref{eq_Pv}, and the lowest value of $\Psi_{\rm{c}}$ satisfying $p_{\rm{v}}=1$ is $\Psi_{\rm{c,min}}$. It can be seen that the analytical results exactly match the simulation results. When  $N_{\rm{PD}}\leq 6$, with the increase of $N_{\rm{PD}}$, $\Psi_{\rm{c,min}}$ decreases and the PR achieves smaller $\Psi_{\rm{c,min}}$ than the TPR. With the further increase of $N_{\rm{PD}}$ from 6 to 10, the lower bound of FOV, $\Psi_{\rm{c,min}}$, for the PR becomes fixed due to \eqref{eq_constraint1_PR} while the $\Psi_{\rm{c,min}}$ for the TPR still decreases and is lower than the $\Psi_{\rm{c,min}}$ for the PR. For $N_{\rm{PD}}>10$, $\Psi_{\rm{c,min}}$ does not change anymore for TPR as well due to \eqref{eq_constraint1_TPR}. It is noted that the Monte-Carlo simulations are also performed for UEs with the random orientation model proposed in \cite{MDSorientation} and the results are matched with the analytical derivation for vertical-orientated UEs as well.

\begin{figure}[!t]
	\centering
	\begin{subfigure}[h]{0.45\textwidth}
		\centering
		\includegraphics[width=1\columnwidth]{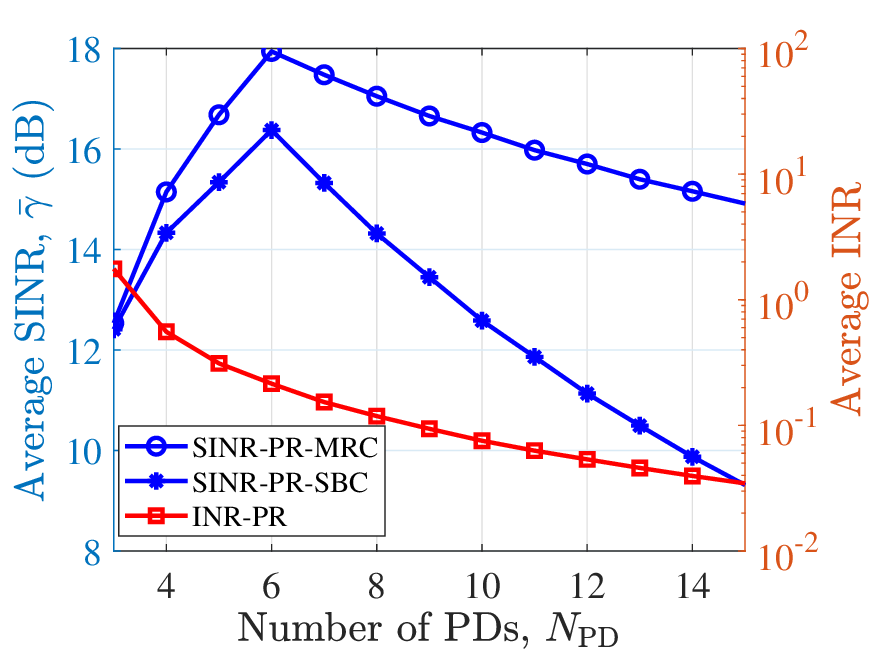}
		\caption{$N_0=10^{-20}$ A$^2$/Hz}
		\label{fig_SINRandINR_N20}
	\end{subfigure}%
	\\
	\begin{subfigure}[h]{0.45\textwidth}
		\centering
		\includegraphics[width=1\columnwidth]{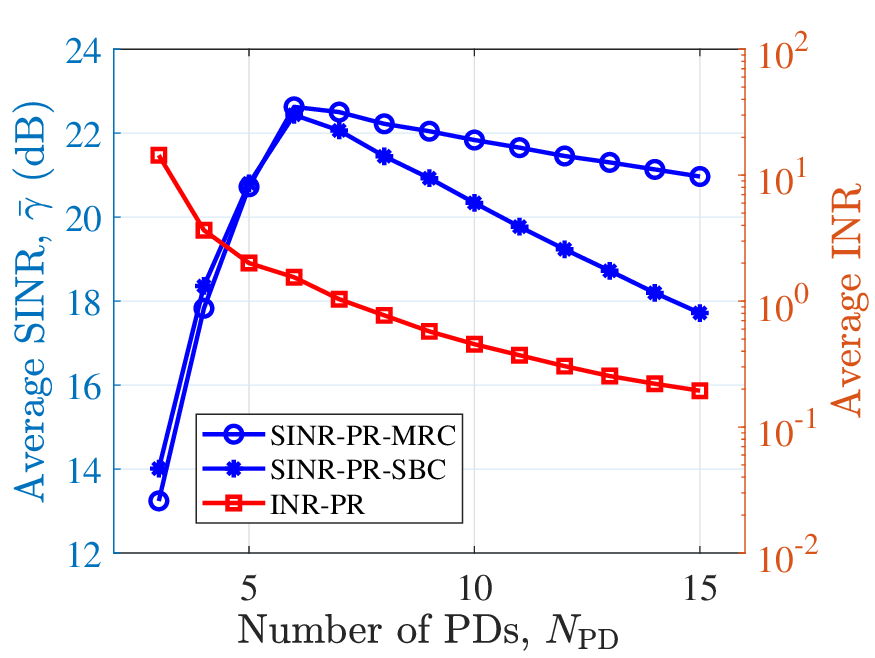}
		\caption{$N_0=10^{-21}$ A$^2$/Hz}
		\label{fig_SINRandINR_N21}
	\end{subfigure}
	\\
	\begin{subfigure}[h]{0.45\textwidth}
		\centering
		\includegraphics[width=1\columnwidth]{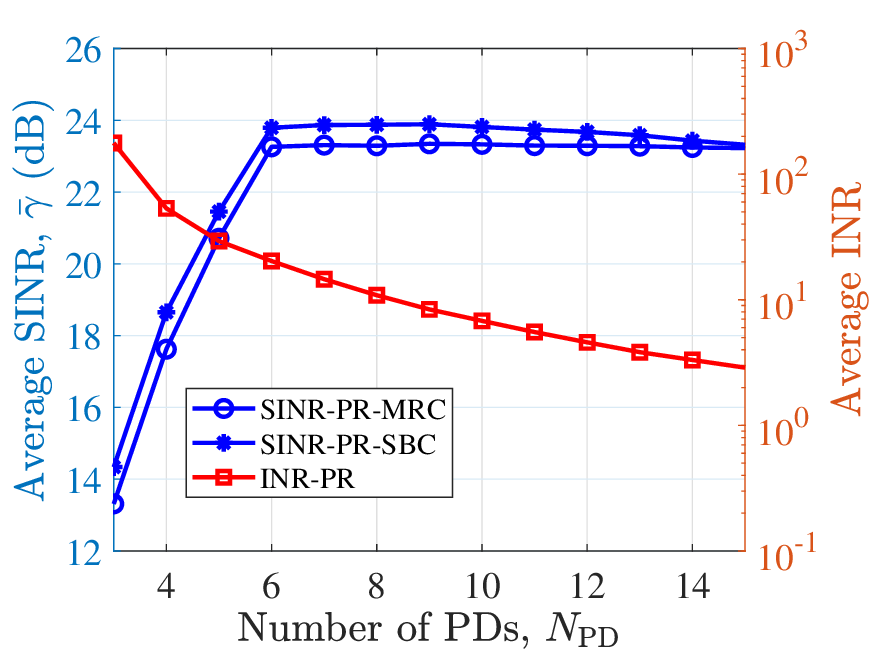}
		\caption{ $N_0=10^{-22}$ A$^2$/Hz}
		\label{fig_SINRandINR_N22}
	\end{subfigure}
	\caption{The performance  comparison between MRC and SBC for the SS (PR, $B_{\rm{t}}$=100 MHz).}
	\label{fig_MRC_vs_SBC}
		\vspace{-10pt}
\end{figure}

2) \textit{MRC vs. SBC}: The MRC scheme is known to achieve better performance than the SBC scheme when there is no interference in the system. However, this may not be true when the interference is taken into consideration. To demonstrate the performance comparison between the MRC and SBC scheme, the simulation is carried out for different noise levels. In Fig. \ref{fig_SINRandINR_N20}, the noise power spectrum density level $N_0$ is $10^{-20}$ A$^2$/Hz. When $N_{\rm{PD}}=3$, the level of interference is slightly higher than the noise, and the MRC schemes achieves similar average SINR as the SBC scheme. With the increase of $N_{\rm{PD}}$, the noise level becomes higher than the interference level and the system gradually becomes noise dominated. It can be seen that in a noise-limited system, MRC outperforms SBC in terms of average SINR. The noise power spectrum density level $N_0$ is $10^{-21}$ A$^2$/Hz in Fig. \ref{fig_SINRandINR_N21}. When $N_{\rm{PD}}\leq 6$, the interference level is higher than the noise level and SBC performs slightly better than MRC. However, the noise starts to rise above the interference level with the further increase of $N_{\rm{PD}}$ and thus MRC outperforms SBC. For a noise power spectrum density level of $10^{-22}$ A$^2$/Hz, Fig. \ref{fig_SINRandINR_N22} depicts an interference-limited system and the SBC scheme achieves a higher average SINR than the MRC scheme for all values of $N_{\rm{PD}}$. In brief, when the system is interference dominated, SBC is a better combining scheme. Otherwise, MRC outperforms SBC when considering the average SINR.
\begin{figure}[t!]
	\centering
	\begin{subfigure}[h]{0.45\textwidth}
		\centering
		\includegraphics[width=\columnwidth]{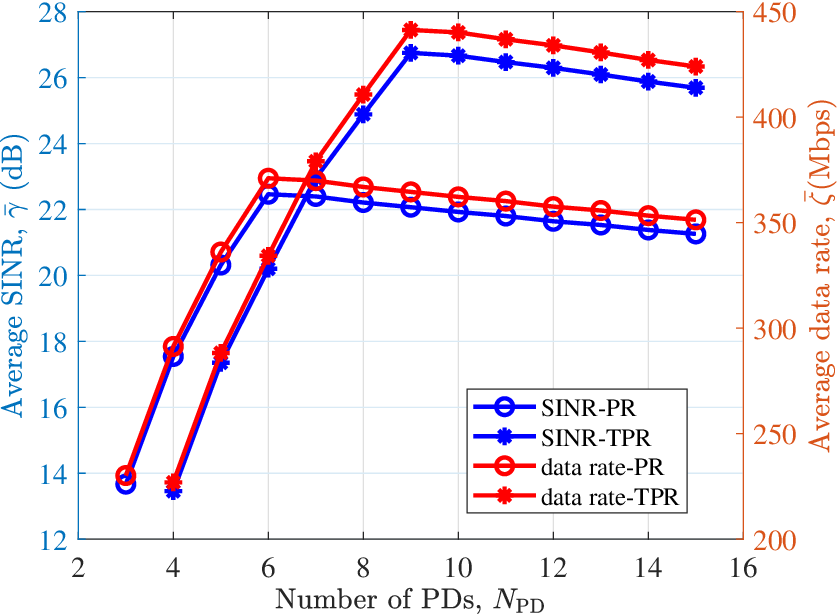}
		\caption{$B_{\rm{t}}$=100 MHz}
		\label{fig_SINR_vs_DR_100}
	\end{subfigure}%
	\\
	\begin{subfigure}[h]{0.45\textwidth}
		\centering
		\includegraphics[width=\columnwidth]{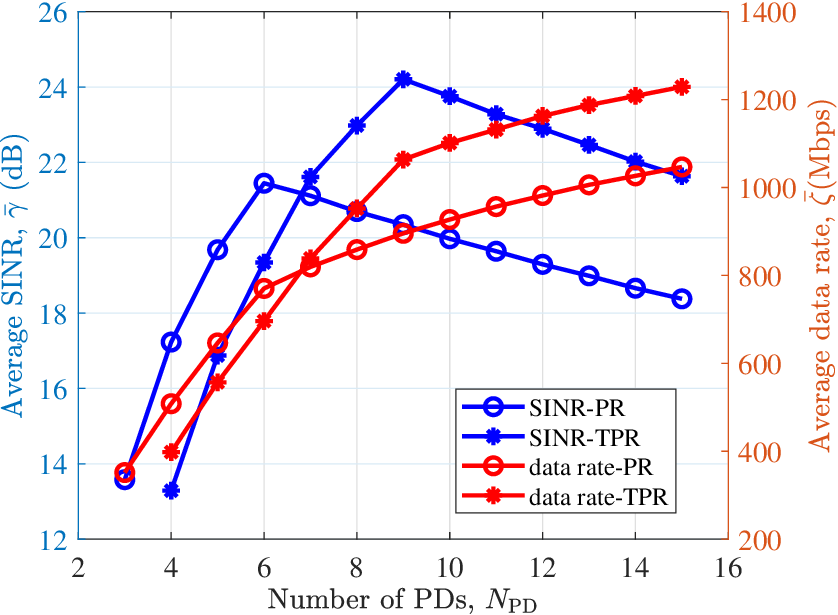}
		\caption{$B_{\rm{t}}$=500 MHz}
		\label{fig_SINR_vs_DR_500}
	\end{subfigure}
	\caption{The performance comparison between PR and TPR configurations considering different transmitter bandwidth in the SS system (MRC, $N_0=10^{-21}$ A$^2$/Hz).}
	\label{fig_SINR_vs_DR}
		\vspace{-10pt}
\end{figure} 

3) \textit{PR vs. TPR}: Fig. \ref{fig_SINR_vs_DR} manifests how the number of PDs $N_{\rm{PD}}$  affects the system performance for both PR and TPR configurations. The average SINR in Fig. \ref{fig_SINR_vs_DR_100} and Fig. \ref{fig_SINR_vs_DR_500} exhibit the same tendency. For  fair comparisons, it is assumed that the total physical area, $A_{\rm{t}}=N_{\rm{PD}}A_{\rm{p}}$, of the ADRs  should be the same. Hence, the increase in $N_{\rm{PD}}$ will lead to the decrease in the physical area $A_{\rm{p}}$ on each PD, which means less received power. Nevertheless, when $N_{\rm{PD}}$ increases from 3 to 6, the average SINR for the PR and TPR both increase. The increase is caused by the decrease in $\Psi_{\rm{c,min}}$ as displayed in Fig. \ref{fig_FOV_vs_N}, which leads to a higher channel gain and compensates for the power loss due to the decrease in $A_{\rm{p}}$. In terms of the PR, the further growth of $N_{\rm{PD}}$ leads to the decline in the average SINR since $\Psi_{\rm{c,min}}$ does not change anymore. In comparison, the average SINR for the TPR increases until $N_{\rm{PD}}=9$ as $\Psi_{\rm{c,min}}$ is still decreasing. When $N_{\rm{PD}}$ increases from 9 to 10 for the TPR, $\Psi_{\rm{c,min}}$ decreases slightly from $21^\circ$ to $20^\circ$ as demonstrated in Fig. \ref{fig_FOV_vs_N}. However, the power loss caused by the reduction in  $A_{\rm{p}}$ exceeds the increase of received power gained from the small decrease in $\Psi_{\rm{c,min}}$, and thus the average SINR drops. Considering $N_{\rm{PD}}\geq10$,  $\Psi_{\rm{c,min}}$ is fixed and the average SINR declines as $A_{\rm{p}}$ reduces. Hence, in terms of the average SINR, the optimum values of $N_{\rm{PD}}$ are 6 and 9 for PR and TPR, respectively. It can also be observed that the PR outperforms the TPR with regard to both the average SINR and average data rate when $N_{\rm{PD}}\leq 6$ since the PR has smaller $\Psi_{\rm{c,min}}$. On the other hand, when $N_{\rm{PD}}>6$, the TPR has smaller $\Psi_{\rm{c,min}}$ than the PR and hence achieves better performance.
\begin{figure}[!t]
	\centering
	\includegraphics[width=\columnwidth]{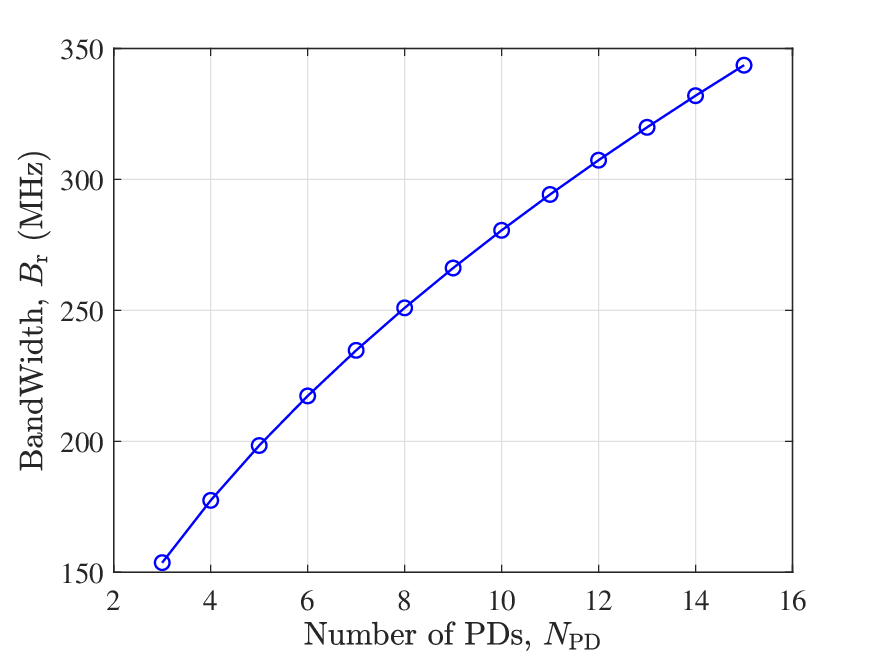}
	\caption{ Receiver bandwidth versus the number of PDs, $N_{\rm{PD}}$.}
	\label{fig_Br_fov}
	  	\vspace{-10pt}
\end{figure}

4) \textit{Receiver bandwidth vs. transmitter bandwidth}: Based on \eqref{eq_br}, the relationship between the receiver bandwidth, $B_{\rm{r}}$, and the number of PDs, $N_{\rm{PD}}$, is manifested in \mbox{Fig. \ref{fig_Br_fov}}. When $N_{\rm{PD}}$ increases from 3 to 15, the bandwidth increases from 150 MHz to around 350 MHz due to the decrease in the physical area of each PD. The communication bandwidth $B_{\rm{L}}$ is the minimum value between the receiver bandwidth and the transmitter bandwidth, which is denoted as $B_{\rm{L}}=\min(B_{\rm{r}},B_{\rm{t}})$. 
The user data rate is determined by the SINR and the communication bandwidth. When the transmitter bandwidth $B_{\rm{t}}$ is 100 MHz, which is less than $B_{\rm{r}}$ for all values of $N_{\rm{PD}}$ in Fig. \ref{fig_Br_fov}, the communication bandwidth is limited by the transmitter and thus $B_{\rm{L}}=100$ MHz. As $B_{\rm{L}}$ does not vary according to $N_{\rm{PD}}$, the average data rate follows the same trend as the average SINR in Fig. \ref{fig_SINR_vs_DR_100}.  By increasing the transmitter bandwidth $B_{\rm{t}}$ to 500 MHz,  $B_{\rm{r}} \leq B_{\rm{t}}$ for all $N_{\rm{PD}}$ and the communication bandwidth is limited by the receiver side. Hence, the communication bandwidth $B_{\rm{L}}=B_{\rm{r}}$. In Fig. \ref{fig_SINR_vs_DR_500}, with the increase of $N_{\rm{PD}}$, the average SINR first increases and then decreases, which peaks at $N_{\rm{PD}}=6$ and $N_{\rm{PD}}=9$ for the PR and TPR, respectively. In contrast, when $N_{\rm{PD}}$ grows, the average data rate increases even when the SINR degrades, which is due to the greater bandwidth. 
To sum up, the PR with $N_{\rm{PD}}=6$ and TPR with $N_{\rm{PD}}=9$ achieve the highest average data rate for transmitter-bandwidth-limited systems whereas the average data rate peaks at $N_{\rm{PD}}=15$ in a receiver-bandwidth-limited system for both PR and TPR. 

\begin{figure}[!t]
	\centering
	\includegraphics[width=\columnwidth]{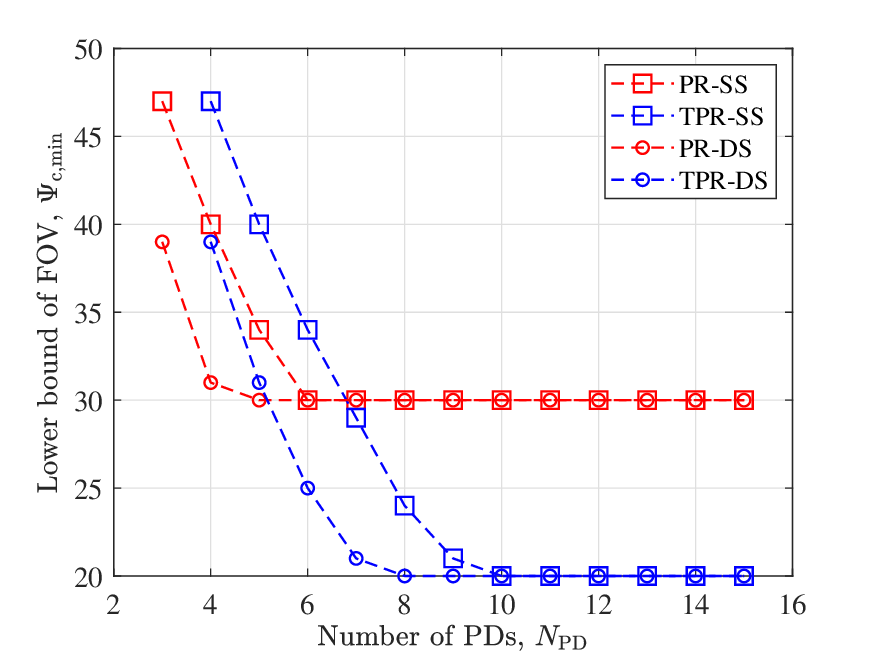}
	\caption{ The relationship between $\Psi_{\rm{c,min}}$ and the number of PDs, $N_{\rm{PD}}$, on an ADR for the DS system.}
	\label{fig_FOV_vs_NumPD_DS}
		  	\vspace{-10pt}
\end{figure}
\subsection{Performance Analysis for DS cells}
1) \textit{Lower bound of FOV}: Fig. \ref{fig_FOV_vs_NumPD_DS} demonstrates the change in the minimum FOV against the number of PDs for the DS configuration. As shown for the PR, with the increase in $N_{\rm{PD}}$, the lower bound of FOV $\Psi_{\rm{c,min}}$ for the DS cells decreases and converges to $30^\circ$  at $N_{\rm{PD}}= 5$, which is earlier than the SS configuration. When $N_{\rm{PD}}\leq 5$, the $\Psi_{\rm{c,min}}$ for the DS configuration is lower than the counterpart in the SS configuration. In terms of the TPR, the minimum FOV for DS converges to $20^\circ$ at $N_{\rm{PD}}= 8$ and has smaller $\Psi_{\rm{c,min}}$ than the SS when $N_{\rm{PD}}\leq 9$.

\begin{figure}[!t]
	\centering
	\includegraphics[width=\columnwidth]{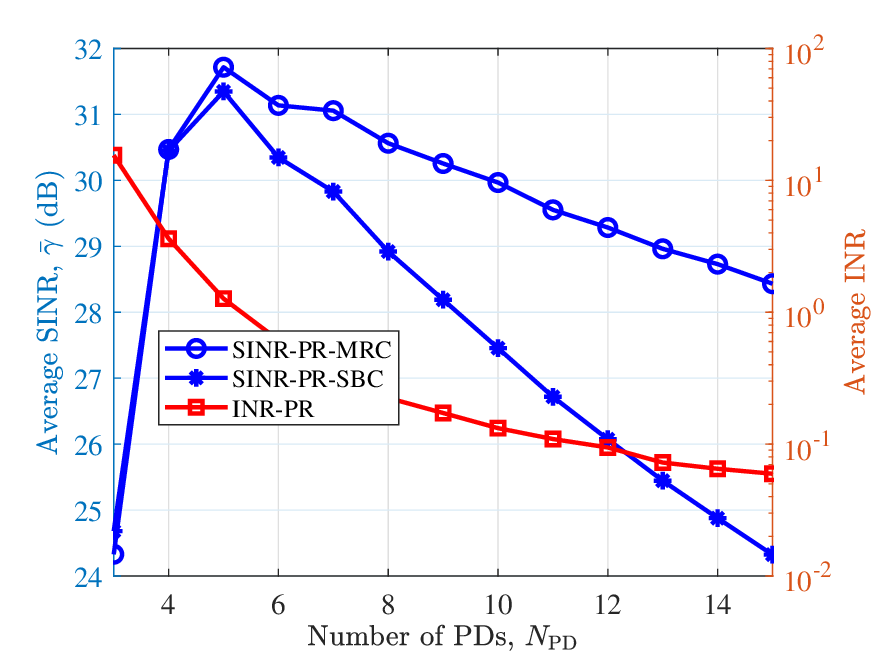}
	\caption{ The  performance comparison between MRC and SBC in DS cells (PR, $N_0=10^{-22}$ A$^2$/Hz, $B_{\rm{t}}$=100 MHz).}
	\label{fig_SINRandINR_N20_DS}
		  	\vspace{-10pt}
\end{figure}

2) \textit{MRC vs. SBC}: Previously, in Fig. \ref{fig_SINRandINR_N22}, we have shown the performance comparison between the MRC and SBC for the SS system with $N_0=10^{-22}$ A$^2$/Hz. Due to the low level of noise power spectral density, the system is mostly interference-limited and SBC outperforms MRC for all given values of $N_{\rm{PD}}$. By adopting the DS configuration, as shown in Fig. \ref{fig_SINRandINR_N20_DS}, the average interference to noise ratio (INR) is larger than 1 only when there are 3 or 4 PDs on the PR. Whereas for $N_{\rm{PD}}\geq 5$, the noise plays a similar or more important role than the interference. Therefore, compared with the SS system, for the same level of noise power spectral density, the average INR of the DS system degrades substantially. This indicates that the DS system can mitigate interference. With regards to the average SINR, MRC and SBC have similar performance when $N_{\rm{PD}}\leq4$ whereas MRC outperforms SBC for $N_{\rm{PD}}>4$. From the above analyses, it can be deducted that MRC is a better combining scheme for the DS system with respect to the three different levels of $N_0$ given previously in this study.

\begin{figure}[!t]
	\centering
	\includegraphics[width=0.8\columnwidth]{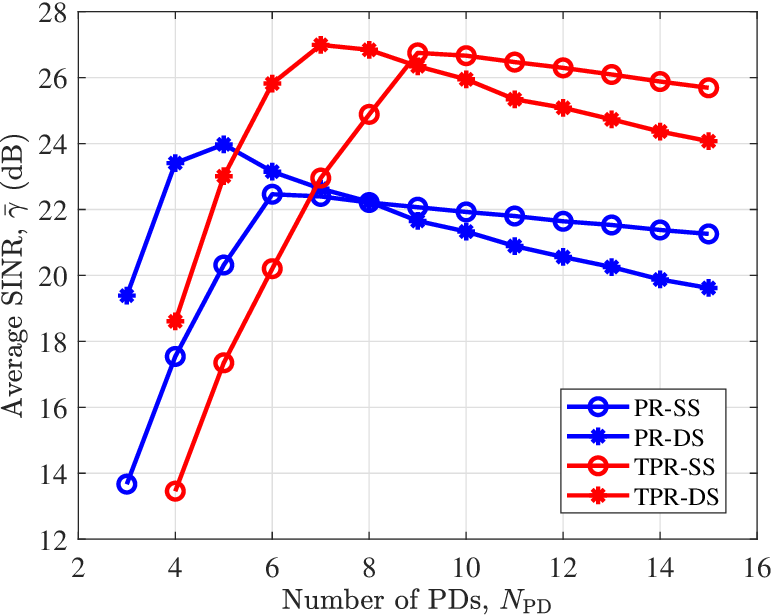}
	\caption{ The performance comparison between PRs and TPRs in DS cells (MRC, $N_0=10^{-21}$ A$^2$/Hz).}
	\label{fig_SINR_NumPD_multi_N21}
	\vspace{-10pt}
\end{figure}

3) \textit{PR vs. TPR}: Fig. \ref{fig_SINR_NumPD_multi_N21} illustrates the performance comparison between PR and TPR configurations. With respect to the PR, when $N_{\rm{PD}}$ rises from 3 to 5, the average SINR increases from 19.3 dB to 24 dB, where the increment comes from the decline in $\Psi_{\rm{c,min}}$. For the same reason, with regards to the TPR, the average SINR grows from 18.5 dB to 27 dB when $N_{\rm{PD}}$ increases from 4 to 7. The PR and TPR achieves the peak at $N_{\rm{PD}}=5$ and $N_{\rm{PD}}=7$ respectively. For both PR and TPR configurations, after the peak points, the average SINR drops due to the reduction in the 
area $A_{\rm{p}}$ of each PD, which leads to less physical power. With regard to TPR, changing $N_{\rm{PD}}$ from 7 to 8 results in the decrease in $\Psi_{\rm{c,min}}$, which leads to a channel gain boost. However, the gain cannot compensate for the power loss stems that from the reduction in $A_{\rm{p}}$ and thus the system performance degrades. When $N_{\rm{PD}}\leq 5$, the PR outperforms the TPR, otherwise, the TPR is a preferred structure. In conclusion, the optimum number of PDs is 5 and 7 for PRs and TPRs, respectively. In addition, the TPR with  $N_{\rm{PD}}=7$ outperforms the PR with $N_{\rm{PD}}=5$ in terms of the average SINR. 

\begin{figure}[t!]
	\centering
	\includegraphics[width=0.9\columnwidth]{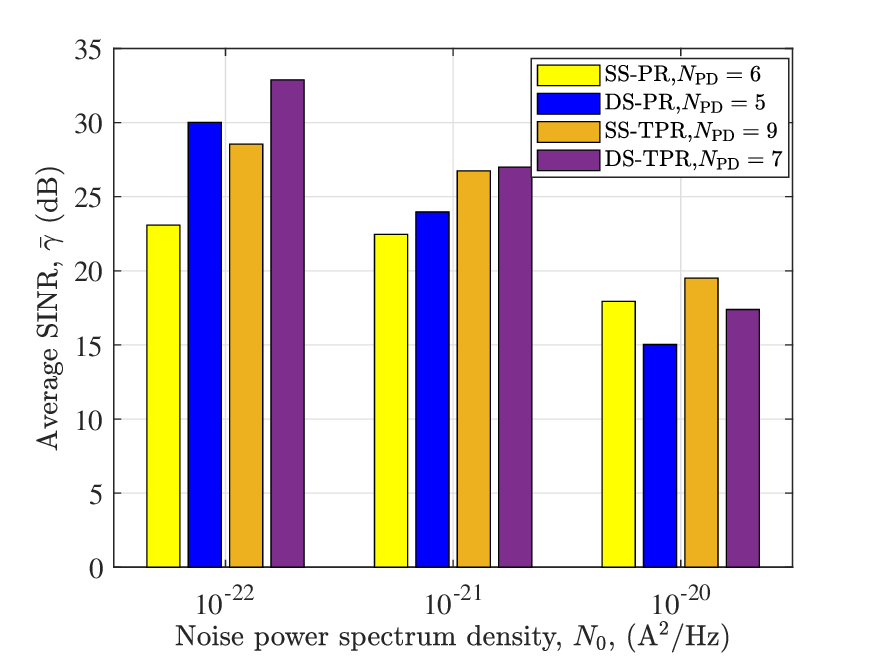}
	\caption{ The performance comparison between the DS and SS systems.}
	\label{fig_SINR_vs_N0}
	\vspace{-10pt}
\end{figure}

4) \textit{DS vs. SS}: Fig. \ref{fig_SINR_vs_N0} demonstrates the performance comparison between the DS system and the SS system with respect to different levels of noise power spectrum density, $N_0$. The typical value of $N_0$ for a PD is $10^{-21}$ A$^2$/Hz, at which the DS system achieves a slightly higher average SINR than the SS system for both PRs and TPRs. By reducing the noise power spectrum density level to  $10^{-22}$ A$^2$/Hz, the system becomes interference-limited. With the aid of the DS configuration, the ADR can suppress the signal power from interfering APs by attenuating the NLOS path. Hence, when $N_0=10^{-22}$ A$^2$/Hz, the average SINR of DS cells is 7 dB and 5 dB higher than the average SINR of SS cells for PRs and TPRs, respectively. For a noise spectrum density level of $10^{-20}$ A$^2$/Hz, the system becomes noise limited, in which the DS could not improve system performance by reducing the power of interference signal. In addition, the received power is halved when the DS configuration is applied, and thus the SS outperforms the DS in terms of the average SINR.

\section{Conclusions}	
This paper investigates the ICI mitigation in LiFi networks using ADRs. The impact of the diffuse link considering random UE orientation is studied and it is shown that both LOS and diffuse links have an important influence on the system performance. The performance of different ADR structures are compared and the optimized ADR structure is proposed for the considered scenario, where the method can be extended to other scenarios easily. By studying systems with different levels of noise power spectrum density, we showed that when the system is noise-limited,  MRC outperforms  SBC, otherwise, SBC is the preferred combining scheme. In an interference-limited system or noise-plus-interference limited system, the adoption of the DS cell configuration can further mitigate the NLOS interference and thus improve the system performance. However, the limitation of the DS cell is that the transmit power is equally split to the positive and negative sources, which degrades the performance of the noise-limited system.

\begin{appendices}
	
	\section{Proof of \eqref{eq_ellipse}}
	\label{appendix1}
	The visible area of the 1-st PD with $\omega^{1}_{\rm{PD}}=0$ is illustrated in Fig. \ref{fig_3d_coverage}a. The point \textbf{O} represents the location of the UE, $\mathbf{p}_{\rm{UE}}$, and $\mathbf{OE}$ represents the normal vector of the PD. The intersection point of the normal vector with the ceiling is $\mathbf{E}$. The coordinate of $\mathbf{E}$ is denoted as  $(x_{\rm{e}}, y_{\rm{e}}, z_{\rm{e}})$ and points $\mathbf{A}$, $\mathbf{B}$, $\mathbf{C}$, $\mathbf{D}$, $\mathbf{F}$ are denoted in the same way.   Points $\mathbf{A}$ and $\mathbf{B}$ are the vertices of the ellipse. Points $\mathbf{C}$ and $\mathbf{D}$ are the co-vertices of the ellipse. The centre point of the ellipse is denoted as $\mathbf{F}$. The angle between each of the four vectors, $\mathbf{OA}$, $\mathbf{OB}$, $\mathbf{OC}$, $\mathbf{OD}$ with $\mathbf{OE}$ is $\Psi_{\rm{c}}$. The length of the semi-major and semi-minor axes of the ellipse are represented by $a$ and $b$ separately. The length of semi-major axes $a$ is denoted as: 
	\begin{equation}
		\setcounter{equation}{47}
		\begin{split}
			a=\frac{|\mathbf{AB}|}{2}
			&=\frac{h\sin(2\Psi_{\rm{c}})}{\cos (2\Psi_{\rm{c}})+\cos (2\Theta_{\rm{PD}})}.
		\end{split}
	\end{equation}
	
	As $\mathbf{A}$, $\mathbf{B}$, $\mathbf{E}$, $\mathbf{F}$ and $\mathbf{O}$ are on the same $xz$-plane, \mbox{$y_{\rm{a}}=y_{\rm{b}}=y_{\rm{e}}=y_{\rm{f}}=y_{\rm{UE}}$}. As $\mathbf{A}$, $\mathbf{B}$, $\mathbf{C}$, $\mathbf{D}$, $\mathbf{E}$ and $\mathbf{F}$ are on the ceiling, \mbox{$z_{\rm{a}}=z_{\rm{b}}=z_{\rm{c}}=z_{\rm{d}}=z_{\rm{f}}=z_{\rm{e}}=z_{\rm{AP}}$}, where $z_{\rm{AP}}$ is the height of the AP. As $\mathbf{C}$ and $\mathbf{D}$ are on the same $yz$-plane with  $\mathbf{F}$, the coordinates $x$ of these points are represented by: 
	\begin{equation}
		\begin{split}
			x_{\rm{c}}&=x_{\rm{d}}=x_{\rm{f}}=x_{\rm{UE}}+(a-h\tan(\Psi_{\rm{c}}-\Theta_{\rm{PD}}))\\
			&=x_{\rm{UE}}+\frac{h\sin(2\Theta_{\rm{PD}})}{\cos (2\Psi_{\rm{c}})+\cos (2\Theta_{\rm{PD}})}.
		\end{split}
	\end{equation}
	From Fig. \ref{fig_3d_coverage}, we know that $x_{\rm{e}}=h\tan(\Theta_{\rm{PD}})+x_{\rm{UE}}$. Based on the parameters above, we have \mbox{$\mathbf{OE}=(h\tan(\theta_{\rm{PD}}), 0, h)$} and \mbox{$\mathbf{OC}=(\frac{h\sin(2\Theta_{\rm{PD}})}{\cos (2\Psi_{\rm{c}})+\cos (2\Theta_{\rm{PD}})}, b, h)$}. Since $\cos(\Psi_{\rm{c}})=\frac{\mathbf{OE}\cdot\mathbf{OC}}{|\mathbf{OE}||\mathbf{OC}|}$, $b$ can be obtained as:
	\begin{equation}
		b=\frac{\sqrt{2}h\sin(\Psi_{\rm{c}})}{\sqrt{\cos (2\Psi_{\rm{c}})+\cos (2\Theta_{\rm{PD}})}}.
	\end{equation}
	Consequently, the equation of the ellipse is thus given by:
	\begin{equation}
		\frac{(x_{\rm{ellipse}}-x_{\rm{f}})^2}{a^2}+\frac{(y_{\rm{ellipse}}-y_{\rm{f}})^2}{b^2}=1.
	\end{equation}  
	The visible area of $p$-th PD can be obtained by rotating the visible area of the 1-st PD around the line $(x=x_{\rm{UE}},y=y_{\rm{UE}})$ with an angle of $\omega_{{\rm{PD}},p}$. 	
	
	\section{Proof of \eqref{eq_Psi_c2_min}}
	\label{appendix2}
	Substituting \eqref{eq_d13} and \eqref{eq_d2_rc_d3} into \eqref{eq_d12}, we can get:
	\begin{equation}
		-d_{\rm{c}}\cos\omega_{\rm{c}}\tan(\Theta_{\rm{PD}})=h-\frac{\sqrt{h^2+d^2_{\rm{c}}}\cos(\Psi_{\rm{c}})}{\cos(\Theta_{\rm{PD}})}.
		\label{eq_derivetive}
	\end{equation}
	Substituting \eqref{eq_Psi_total} into \eqref{eq_derivetive}, the elevation angle of each PD, $\Theta_{\rm{PD}}$, on PRs is derived as:
	\begin{equation}
		\Theta_{\rm{PD}}=F_1(d_{\rm{c}})=\tan^{-1}\frac{f_1(d_{\rm{c}})}{f_2(d_{\rm{c}})}, 
		\label{Theta}
	\end{equation}
	where
	\begin{equation}
		\begin{split}
			&f_1(d_{\rm{c}})=\sqrt{h^2+d^2_{\rm{c}}}\cos(\Psi_{\rm{total}})-h, \\
			&f_2(d_{\rm{c}})={d_{\rm{c}}\cos(\omega_{\rm{c}})-\sqrt{h^2+d^2_{\rm{c}}}\sin(\Psi_{\rm{total}})}.
		\end{split}
	\end{equation}
	The function $F_1$ has one zero at $z_1=h\tan(\Psi_{\rm{total}})$ and one pole at $p_1=\frac{h\sin(\Psi_{\rm{total}})}{\sqrt{\cos^2(\Psi_{\rm{total}})-\sin^2(\omega_{\rm{c}})}}$. The derivative of $F_1(d_{\rm{c}})$ is given by:
	\begin{equation}
		\resizebox{0.9\hsize}{!}{$
			\frac{\partial F_1}{\partial d_{\rm{c}}}=\\ -\frac{h^2\cos(\omega_{\rm{c}})\cos(\Psi_{\rm{total}})-h\sqrt{h^2+d^2_{\rm{c}}}\cos(\omega_{\rm{c}})+hd_{\rm{c}}\sin(\Psi_{\rm{total}})}{(f^2_1(d_{\rm{c}})+f^2_2(d_{\rm{c}}))\sqrt{h^2+d^2_{\rm{c}}}}.
			$}
	\end{equation}
	By calculating the $d_{\rm{c}}$ satisfying $\frac{\partial F_1}{\partial d_{\rm{c}}}=0$, the two roots are denoted as:
	\begin{equation}
		d_{\rm{c1}}
		=\frac{h\cos(\omega_{\rm{c}})\sin(\Psi_{\rm{total}})}{\cos(\Psi_{\rm{total}})-\sin(\omega_{\rm{c}})},~~~
		d_{\rm{c2}}
		=\frac{h\cos(\omega_{\rm{c}})\sin(\Psi_{\rm{total}})}{\cos(\Psi_{\rm{total}})+\sin(\omega_{\rm{c}})}.
	\end{equation}
	When $\sin(\omega_{\rm{c}}) < \cos(\Psi_{\rm{total}})$, it can be proven that  \mbox{$0 < d_{\rm{c2}} < z_1 =h\tan(\Psi_{\rm{total}}) < d_{\rm{c1}}$} and \mbox{$z_1\leq p_1$}. In addition, $\frac{\partial^2 F_1(d_{\rm{c2}})}{\partial^2 d_{\rm{c}}}<0$. Therefore, for $d_{\rm{c}} \in (0,h\tan(\Psi_{\rm{total}})]$, $F_1$ has a local maximum at $d_{\rm{c2}}$.
	When $\sin(\omega_{\rm{c}}) \geq \cos(\Psi_{\rm{total}})$, $p_1$ and $d_{\rm{c1}}$ do not have real value. It can also be proved that $0 < d_{\rm{c2}} < z_1$ and $\frac{\partial^2 F_1(d_{\rm{c2}})}{\partial^2 d_{\rm{c}}}<0$. Similarly, for $d_{\rm{c}} \in (0,h\tan(\Psi_{\rm{total}})]$, there is a local maximum at $d_{\rm{c2}}$. In summary, the upper bound of $F_1$ is thus given by:
	\begin{equation}
		F_{\rm{1,max}}= 
		\begin{dcases}
			F_1(d_{\rm{c2}}),& \text{if} \ d_{\rm{c,min}} \leq d_{\rm{c2}} \\
			F_1(d_{\rm{c,min}}),              & \text{otherwise}.
		\end{dcases}
		\label{eq_ub_F1}
	\end{equation}
	According to \eqref{eq_Psi_total}, $\Psi_{\rm{c}}$ is given by:
	\begin{equation}
		\Psi_{\rm{c}}=F_2(d_{\rm{c}})=\Psi_{\rm{total}}-F_1(d_{\rm{c}}). 
		\label{Psi}
	\end{equation}
	Hence, the lower bound of $F_2$ is denoted as:
	\begin{equation}
		F_{\rm{2,min}}= 
		\begin{dcases}
			F_2(d_{\rm{c2}}),& \text{if} \ d_{\rm{c,min}} \leq d_{\rm{c2}} \\
			F_2(d_{\rm{c,min}}),              & \text{otherwise}.
		\end{dcases}
	\end{equation}
\end{appendices}

\bibliography{report}

\end{document}